\documentclass[prb,twocolumn]{revtex4}

\usepackage{amssymb}
\usepackage{amsmath}
\usepackage{graphicx}

\begin{document}

\author{Satoru Odashima}
\email[E-mail: ]{odashima@sa.infn.it} \homepage[Group Homepage:
]{http://scs.sa.infn.it}
\author{Adolfo Avella}
\author{Ferdinando Mancini}

\affiliation{Dipartimento di Fisica ''E.R. Caianiello" - Laboratorio
Regionale SuperMat, INFM \\ Universit\`{a} degli Studi di Salerno,
I-84081 Baronissi (SA), Italy}

\title{High-order correlation effects in the two-dimensional Hubbard model}

\begin{abstract}
The electronic states of the two-dimensional Hubbard model are
investigated by means of a 4-pole approximation within the Composite
Operator Method. In addition to the conventional Hubbard operators,
we consider other two operators which come from the hierarchy of the
equations of motion and carry information regarding nearest-neighbor
spin and charge configurations. By means of this operatorial basis,
we can study the physics related to the energy scale of $J=4t^2/U$
in addition to the one of $U$. Present results show relevant
physical features, well beyond those previously obtained by means of
a 2-pole approximation, such as a four-band structure with shadow
bands and a quasi-particle peak at the Fermi level. The Fermi level
stays pinned to the band flatness located at ($\pi$,$0$)-point
within a wide range of hole-doping ($0 \leq \delta \leq 0.15$). A
comprehensive analysis of double occupancy, internal energy,
specific heat and entropy features have been also performed. All
reported results are in excellent agreement with the data of
numerical simulations.
\end{abstract}

\date{\today}

\maketitle

\section{Introduction}

The discovery of high-$T_c$ superconductors promoted a largely
diffused revival of interest in strongly correlated electron systems
and fostered the study of many other transition-metal
oxides.\cite{Imada_98} Since the very beginning, the Hubbard model
\cite{Hubbard} has received particular attention as it is retained
a prototype for strongly correlated electron systems and a minimal
model to describe transition-metal oxides. In spite of the deceiving
simplicity of its Hamiltonian, a deep comprehension of all its
physical features is still missing. In particular, it is very
difficult to properly describe the competition between kinetic,
diagonal in momentum space, and potential, diagonal in direct space,
energy terms. The main gross feature of the model is the splitting
of the electronic band into two subbands divided by a gap of the
order $U$. Nowadays, it is well-established that this feature can be
understood in terms of well-known Hubbard operators within a 2-pole
approximation. However, within this framework, inter-site
correlations are poorly taken into account while they are
universally recognized as essential to describe low-energy physics
near half filling. For instance, Kampf and Schrieffer pointed out
that, in highly correlated electron systems, antiferromagnetic spin
fluctuations play a fundamental role in understanding features as
pseudogap and shadow bands.\cite{Kampf_90,Kampf_90a} Recently, the
remarkable progress in experimental techniques has made possible to
reveal such low-energy features in high-$T_c$
superconductors.\cite{Damascelli_03} Antiferromagnetic fluctuations
are often included within self-energy in a phenomenological way.
Therefore, it is quite natural to wonder if, being these features
inherent to the model Hamiltonian, it is possible to derive them
microscopically by means of a suitable analytical tool.
Numerical-simulation techniques, as Exact Diagonalization and
quantum Monte Carlo (QMC), positively answer to this
question,\cite{Dagotto_94,Bulut_94a,Preuss_95,Preuss_97} although
they are applicable only to small clusters due to the exponential
increase of the Hilbert space with the system size. Within the
framework of projection methods, in order to describe inter-site
correlations and related spin and charge fluctuations, it is
necessary to take into account high-order operators which carry
information regarding nearest-neighbor correlations. We will move
along this way.

In this paper, we investigate the electronic states of the Hubbard
model by means of the Composite Operator Method (COM)
\cite{Ishihara_94,Mancini_95,Mancini_95a,Mancini_00,Mancini_04}
which has shown to be capable of describing the physics of many
strongly correlated systems. On one hand, we can find many
similarities between COM and the projection operator method
\cite{Mori_65,Roth_69,Roth_69a,Becker_90,Unger_93,Fulde_95,Zhou_91,Mehlig_95}
and the spectral density approach (self-consistent moment
approach).\cite{Nolting_72,Geipel_88,Nolting_89,Rodriguez_98,Rodriguez_98a}
On the other hand, COM differs from these methods as regards the
conscious exploitation of the presence of unknown parameters in the
theory in order to put constraints on the representation where the
Green's functions are realized. The Hubbard model has been widely
analyzed by means of COM within the 2-pole
approximation.\cite{Mancini_95,Mancini_95a,Mancini_04} In this
approximation, COM reaches a global agreement with numerical
simulations regarding local and thermodynamic quantities. In order
to go beyond the 2-pole approximation, it is necessary either to
evaluate the dynamical corrections or to introduce high-order
operators in the basis. As regards the former case, a fully momentum
and frequency dependent self-energy has been evaluated by means of:
two-site level operators within the non-crossing
approximation,\cite{Matsumoto_96a,Matsumoto_96,Matsumoto_97} loop
decoupling within both the self-consistent Born approximation
\cite{Plakida_99,Prelovsek_02,Krivenko_04} and the iterative
perturbation theory within the dynamical mean field
theory.\cite{Onoda_03} As regards the latter case, Dorneich
\textit{et al.} have introduced in the operatorial basis, in
addition to the conventional Hubbard operators, two nonlocal
composite fields which describe the local electronic transitions
dressed by the nearest-neighbor spin and charge
fluctuations.\cite{Dorneich_99} They have managed to reproduce the
well-known four band structure in good agreement with the QMC data,
but only at half filling to which their formulation is unfortunately
restricted. According to all this, we have decided to examine the
model by means of a 4-pole approximation within the Composite
Operator Method. In addition to the conventional Hubbard operators,
we have also considered in the basis other two operators coming from
the hierarchy of the equations of motion. Within our formulation, we
can evaluate the evolution of each subband up to the second order in
the equations of motion and we have managed to perform the analysis
with finite doping, which was out of the scope in
Ref.~\onlinecite{Dorneich_99}. Our results present: a four-band
structure, a quasi-particle peak at the Fermi level, shadow bands,
band flatness at ($\pi$, $0$)-point, a Fermi level located around
the ($\pi$, $0$)-point and robust with respect to the hole doping
($\delta \leq 0.15$). A comprehensive study, and comparisons with
numerical simulations present in the literature, has been performed
as regards: density of states, band dispersion, double occupancy,
internal energy, specific heat and entropy.

The manuscript is organized as follows. In the next section
(Sec.~\ref{sec_FORMULATION}), we fix the notation regarding the
Hamiltonian and give the general framework of the Composite Operator
Method. In Sec.~\ref{sec_DOS}, the density of states and the
dispersion relation are computed and discussed, also in comparison
with some quantum Monte Carlo results. A detailed comparison with
numerical simulations for many local and thermodynamic quantities is
given in Sec.~\ref{sec_THERMODYNAMICS}. Section~\ref{sec_SUMMARY}
contains summary and conclusions. Some detailed derivations of
formulas used in Sec.~\ref{sec_DOS} are given in the appendix.

\section{Model and Formulation}\label{sec_FORMULATION}
The two-dimensional grand-canonical Hubbard Hamiltonian $\hat{H}$
reads as
\begin{align}\label{Ham}
\hat{H} &= H - \mu \sum_{\mathbf{i}\sigma} n_{\sigma}(i) \\
H &= \sum_{\mathbf{ij}\sigma} t_{\mathbf{ij}}
c^{\dagger}_{\sigma}(i) c_{\sigma}(j)+ U \sum_{\mathbf{i}}
n_{\uparrow}(i) n_{\downarrow}(i)
\end{align}
where $c^{\dagger}_{\sigma}(i)$ and $c_{\sigma}(i)$ are creation and
annihilation operators, respectively, of electrons with spin
$\sigma$ at the site $\mathbf{i}$ $[i=(\mathbf{i},t)]$.
$n_{\sigma}(i) = c^{\dagger}_{\sigma}(i) c_{\sigma}(i)$. $\mu$ is
the chemical potential. $t_{\mathbf{ij}} = - 4t
\alpha_{\mathbf{ij}}$. $\alpha(\mathbf{k}) =
\mathcal{F}[\alpha_{\mathbf{ij}}] = \frac{1}{2}\left\{\cos(k_x
a)+\cos(k_y a)\right\}$. $a$ is the lattice constant. $\mathcal{F}$
is the Fourier transform. $U$ is the on-site Coulomb repulsion.
Here, we consider nearest-neighbor hopping only. We define the
following operatorial basis
\begin{equation}
\psi_{\sigma}(i)=\left( \begin{array}{c}
\psi_{A\sigma}(i) \\
\psi_{B\sigma}(i)
\end{array}\right) =\left( \begin{array}{c}
\xi_{\sigma}(i) \\
\eta_{\sigma}(i) \\
\hline
\xi_{s\sigma}(i) \\
\eta_{s\sigma}(i)
\end{array}\right)
\end{equation}
where $\xi_{\sigma}(i) = \left(1-n_{-\sigma}(i)\right)
c_{\sigma}(i)$ and $\eta_{\sigma}(i) = n_{-\sigma}(i) c_{\sigma}(i)$
are the usual Hubbard operators that describe the transitions of the
local electron number $n(i)=0 \leftrightarrow 1$ and $1
\leftrightarrow 2$, respectively. The equations of motion of the
components of $\psi_{A}(i)$ read as
\begin{align}
& \mathrm{i} \frac{\partial}{\partial t}\xi_{\sigma}(i) = -\mu\xi_{\sigma}(i)
 -4t\left\{ c^{\alpha}_{\sigma}(i)+\pi_{\sigma}(i) \right\} \label{em01a} \\
& \mathrm{i} \frac{\partial}{\partial t}\eta_{\sigma}(i) = (-\mu+U)\eta_{\sigma}(i)
 +4t\pi_{\sigma}(i) \label{em01b}
\end{align}
where
\begin{align}
\pi_{\sigma}(i)&=-n_{-\sigma}(i)c^{\alpha}_{\sigma}(i)
+c^{\dagger}_{-\sigma}(i)c_{\sigma}(i)c^{\alpha}_{-\sigma}(i)
\nonumber \\
&+c_{\sigma}(i)c^{\alpha\dagger}_{-\sigma}(i)c_{-\sigma}(i)\\
c^{\alpha}_{\sigma}(i)&=\sum_{\bf j}\alpha_{\bf ij}c_{\sigma}({\bf j},t).
\end{align}
Now, we can divide $\pi_{\sigma}(i)$ into two operators,
$\pi_{\sigma}(i)=\xi_{s\sigma}(i)+\eta_{s\sigma}(i)$, similarly to
what we have done with $c_{\sigma}(i)$,
$c_{\sigma}(i)=\xi_{\sigma}(i)+\eta_{\sigma}(i)$. That is, we choose
the components of $\psi_{B}(i)$, $\xi_{s\sigma}(i)$ and
$\eta_{s\sigma}(i)$, among the eigenoperators of the two-site
Hubbard model \cite{Avella_01} and of the local interaction term of
the Hamiltonian (\ref{Ham}). Then, they read as
\begin{align}
\xi_{s\sigma}(i)&=-n_{-\sigma}(i)\xi^{\alpha}_{\sigma}(i) \nonumber \\
&+c^{\dagger}_{-\sigma}(i)c_{\sigma}(i)\xi^{\alpha}_{-\sigma}(i)
 +c_{\sigma}(i)\eta^{\alpha\dagger}_{-\sigma}(i)c_{-\sigma}(i) \\
\eta_{s\sigma}(i)&=-n_{-\sigma}(i)\eta^{\alpha}_{\sigma}(i) \nonumber \\
&+c^{\dagger}_{-\sigma}(i)c_{\sigma}(i)\eta^{\alpha}_{-\sigma}(i)
 +c_{\sigma}(i)\xi^{\alpha\dagger}_{-\sigma}(i)c_{-\sigma}(i).
\end{align}
It is clear now that $\psi_{B}(i)$ describes nearest-neighbor
composite excitations and carries information regarding surrounding
spin and charge configurations.\cite{Odashima_04} According to the
way we have chosen them,\cite{Avella_01} $\xi_{\sigma}(i)$ and
$\xi_{s\sigma}(i)$ belong to the energy class of the lower Hubbard
subband and $\eta_{\sigma}(i)$ and $\eta_{s\sigma}(i)$ belong to the
energy class of the upper Hubbard subband (see Eqs.~(\ref{em01a}),
(\ref{em01b}), (\ref{em02a}) and (\ref{em02b})).

Within the Composite Operator Method, once we choose a $n$-component
operatorial basis $\psi$, its equation of motion reads as
\begin{equation}
\mathrm{i}\frac{\partial}{\partial t}\psi(i)=\sum_{\bf j}\epsilon({\bf i},{\bf j})
 \psi({\bf j},t)+\delta j(i)
\end{equation}
where $\epsilon({\bf k})=m({\bf k})I^{-1}({\bf k})$ is a $n \times n$ matrix with
\begin{align}
&I({\bf k})=\mathcal{F}\langle \{ \psi({\bf i},t),
 \psi^{\dagger}({\bf j},t) \} \rangle\\
&m({\bf k})=\mathcal{F}\langle \{ i\frac{\partial}{\partial t}\psi({\bf i},t),
 \psi^{\dagger}({\bf j},t) \} \rangle.
\end{align}
The spinor notation is understood and $\langle \cdots \rangle$
denotes the thermal average in the grand-canonical ensemble. The
expression of $\epsilon({\bf k})$ comes from the request
$\langle \{ \delta j({\bf i},t), \psi^{\dagger}({\bf j},t) \} \rangle =0$.
This constraint assures that the residual term $\delta j(i)$ contains
only the dynamical corrections in terms of orthogonal components to
the chosen basis. Neglecting $\delta j(i)$ gives the $n$-pole
approximation for the retarded electronic Green's function
$G(\omega, {\bf k})=\mathcal{F}\left\langle
R[\psi(i)\psi^\dagger(j)] \right\rangle $:
\begin{equation}
G(\omega, {\bf k})=\sum_{i=1}^n\frac{\sigma_{i}({\bf k})}{\omega
-E_{i}({\bf k})+\mathrm{i}\delta}
\end{equation}
where $E_{i}({\bf k})$ are the eigenvalues of $\epsilon({\bf k})$
and $\sigma_{i}({\bf k})$, the spectral functions, can be computed
in terms of the eigenvectors of $\epsilon({\bf k})$ and of the
elements of $I({\bf k})$.\cite{Mancini_00,Mancini_04}

In the last years, the 2-pole approximation, that is,
$\psi=\psi_{A}$, has been analyzed in great detail.\cite{Mancini_04}
In the present paper, we perform a 4-pole analysis by enlarging the
operatorial basis with the introduction of $\psi_{B}$. In this case,
$I({\bf k})$ reads as
\begin{align}
I({\bf k}) & = \left( \begin{array}{cc}
I_{AA}({\bf k}) & I_{AB}({\bf k}) \\
I_{AB}({\bf k}) & I_{BB}({\bf k})
\end{array} \right) \nonumber \\
& = \left( \begin{array}{cc|cc}
I_{11} & 0 & I_{13}({\bf k}) & 0 \\
0 & I_{22} & 0 & I_{24}({\bf k}) \\
\hline
I_{13}({\bf k}) & 0 & I_{33}({\bf k}) & I_{34}({\bf k}) \\
0 & I_{24}({\bf k}) & I_{34}({\bf k}) & I_{44}({\bf k})
\end{array} \right)
\end{align}
where
\begin{align}
& I_{11\sigma}=1-\langle n_{-\sigma}\rangle \\
& I_{22\sigma}=\langle n_{-\sigma}\rangle \\
& I_{13\sigma}({\bf k})=\Delta_{\sigma}-\alpha({\bf k})\left( \langle n_{-\sigma} \rangle -p_{\sigma} \right) \\
& I_{24\sigma}({\bf k})=-\Delta_{\sigma}-\alpha({\bf k})p_{\sigma}
\end{align}
with
\begin{align}
\Delta_{\sigma}&=\langle
\eta^{\dagger}_{-\sigma}\eta^{\alpha}_{-\sigma} \rangle
-\langle \xi^{\dagger}_{-\sigma}\xi^{\alpha}_{-\sigma} \rangle \\
p_{\sigma}&=\langle n_{-\sigma}n^{\alpha}_{-\sigma} \rangle +\langle
c^{\dagger}_{-\sigma}c_{\sigma}(c^{\dagger}_{\sigma}c_{-\sigma})^{\alpha}
\rangle \nonumber \\
&-\langle
c_{\sigma}c_{-\sigma}(c^{\dagger}_{-\sigma}c^{\dagger}_{\sigma})^{\alpha}
\rangle.
\end{align}
The detailed expressions of the elements in the block $I_{BB}({\bf
k})$ are rather complicated and reported in Appendix. In order to
effectively perform calculations, we have decoupled these elements
by paying attention to the particle-hole symmetry enjoined by the
Hamiltonian. Under this transformation, we have
\begin{align}
& \xi_{\sigma}(i) \rightarrow (-1)^{\bf i}\eta^{\dagger}_{\sigma}(i) \\
& \eta_{\sigma}(i) \rightarrow (-1)^{\bf i}\xi^{\dagger}_{\sigma}(i) \\
& \xi_{s\sigma}(i) \rightarrow (-1)^{\bf
i}\eta^{\alpha\dagger}_{\sigma}(i) +(-1)^{\bf i}\eta^{\dagger}_{s\sigma}(i)\\
& \eta_{s\sigma}(i) \rightarrow (-1)^{\bf
i}\xi^{\alpha\dagger}_{\sigma}(i) +(-1)^{\bf
i}\xi^{\dagger}_{s\sigma}(i) .
\end{align}
After these relations, we have the following constraints on the
$I_{\sigma}({\bf i,j})=\langle \{ \psi_{\sigma}({\bf i},t),
 \psi^{\dagger}_{\sigma}({\bf j},t)\} \rangle$ matrix elements
\begin{align}
I_{11}({\bf i,j}) \rightarrow & (-1)^{\bf i+j}I_{22}({\bf j,i}) \\
I_{22}({\bf i,j}) \rightarrow & (-1)^{\bf i+j}I_{11}({\bf j,i}) \\
I_{33}({\bf i,j}) \rightarrow & (-1)^{\bf i+j}\{ I_{22}({\bf
j^{\alpha},i^{\alpha}})+I_{24}({\bf j^{\alpha},i})\}\nonumber \\
+ & (-1)^{\bf i+j}\{I_{42}({\bf j,i^{\alpha}})+I_{44}({\bf j,i}) \} \\
I_{34}({\bf i,j}) \rightarrow & (-1)^{\bf i+j}I_{34}({\bf j,i}) \\
I_{44}({\bf i,j}) \rightarrow & (-1)^{\bf i+j}\{ I_{11}({\bf
j^{\alpha},i^{\alpha}}) +I_{13}({\bf j^{\alpha},i})\}\nonumber \\
+ & (-1)^{\bf i+j}\{I_{31}({\bf j,i^{\alpha}})+I_{33}({\bf j,i}) \}
\end{align}
where ${\bf i}^{\alpha}$ stands for the nearest-neighbor sites of
$\bf i$ (e.g., $I_{13}({\bf j^{\alpha},i})=\sum_{\bf l}\alpha_{\bf
jl}I_{13}({\bf l,i})$). It is worth noticing that our decoupling
procedure exactly satisfies these relations.

Now, we introduce a new operator
\begin{equation}\label{base02}
\bar{\psi}_{B\sigma}(i)=\left( \begin{array}{c}
\bar{\xi}_{s\sigma}(i) \\
\bar{\eta}_{s\sigma}(i)
\end{array}\right)
=\left( \begin{array}{c}
\xi_{s\sigma}(i)-A_{31}(-\mathrm{i}\nabla)\xi_{\sigma}(i) \\
\eta_{s\sigma}(i)-A_{42}(-\mathrm{i}\nabla)\eta_{\sigma}(i)
\end{array}\right)
\end{equation}
with
\begin{align}
& \mathcal{F}\left[ A_{31}(-\mathrm{i}\nabla)\right]=A_{31}({\bf
k})=\frac{I_{13}({\bf k})}{I_{11}} \\
& \mathcal{F}\left[ A_{42}(-\mathrm{i}\nabla)\right]=A_{42}({\bf
k})=\frac{I_{24}({\bf k})}{I_{22}}.
\end{align}
This operator gives $I$ in a block-diagonal form
\begin{align}
I({\bf k})&=\left( \begin{array}{cc}
I_{AA}({\bf k}) & 0 \\
0 & I_{BB}({\bf k})
\end{array}\right)\nonumber \\
&=\left( \begin{array}{cc|cc}
I_{11} & 0 & 0 & 0 \\
0 & I_{22} & 0 & 0 \\
\hline
0 & 0 & \bar{I}_{33}({\bf k}) & \bar{I}_{34}({\bf k}) \\
0 & 0 & \bar{I}_{34}({\bf k}) & \bar{I}_{44}({\bf k})
\end{array}\right)
\end{align}
where
\begin{align}
& \bar{I}_{33}({\bf k})=I_{33}({\bf k})-\frac{I^{2}_{13}({\bf k})}{I_{11}} \\
& \bar{I}_{34}({\bf k})=I_{34}({\bf k}) \\
& \bar{I}_{44}({\bf k})=I_{44}({\bf k})-\frac{I^{2}_{24}({\bf
k})}{I_{22}}.
\end{align}
Hereafter, we will use this new operator $\bar{\psi}_{B\sigma}(i)$
instead of $\psi_{B\sigma}(i)$ as it allows to more easily
distinguish the contributions to single-particle properties of type
$B$ operators from those of type $A$. Then
\begin{equation}
\psi_{\sigma}(i)=\left( \begin{array}{c}
\psi_{A\sigma}(i) \\
\bar{\psi}_{B\sigma}(i)
\end{array} \right).
\end{equation}

\begin{figure}[tbp]
\includegraphics[width=0.43\textwidth]{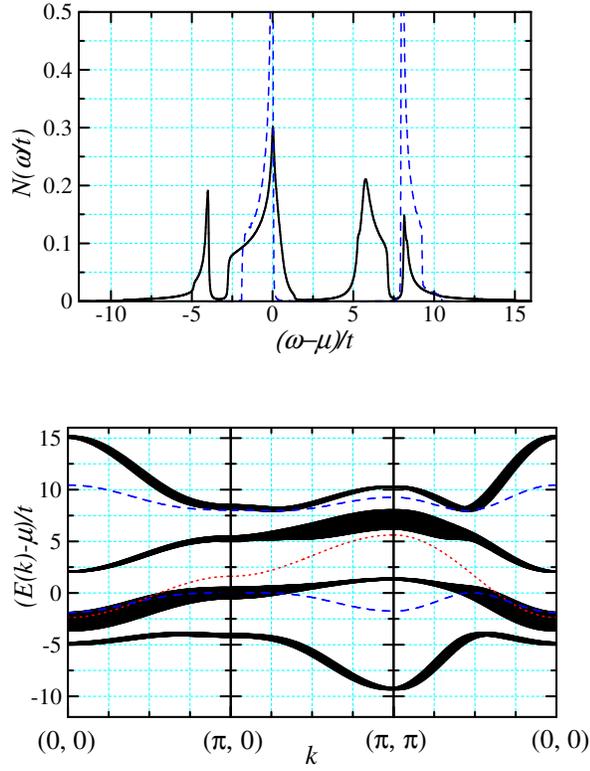}
\caption{The density of states and the corresponding dispersion
relation (solid line) at $U/t=8$, $n=0.90$ and $T/t=0.01$. The width
of the dispersion line represents the intensity of peak. The 2-pole
solution (dashed line) and the non-interacting case (dotted line)
are also reported.}\label{fig01}
\end{figure}

\begin{figure}[tbp]
\includegraphics[width=0.43\textwidth]{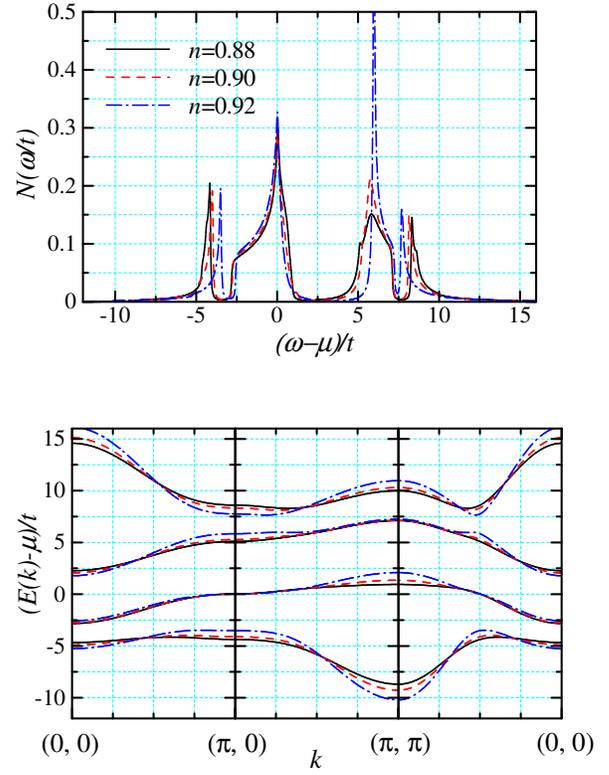}
\caption{Doping dependence of density of states and dispersion
relation at $n=0.88$, $0.90$ and $0.92$. Other external parameters
are the same as in Fig. \ref{fig01}.}\label{fig02}
\end{figure}

By means of this new basis, we have the matrix $\epsilon=mI^{-1}$,
\begin{align}
\epsilon({\bf k})& =\left( \begin{array}{cc}
\epsilon_{AA}({\bf k}) & \epsilon_{AB}({\bf k}) \\
\epsilon_{BA}({\bf k}) & \epsilon_{BB}({\bf k})
\end{array} \right)\nonumber \\
&=\left( \begin{array}{cc|cc}
\epsilon_{11}({\bf k}) & \epsilon_{12}({\bf k}) & \epsilon_{13}({\bf k}) & \epsilon_{14}({\bf k}) \\
\epsilon_{21}({\bf k}) & \epsilon_{22}({\bf k}) & \epsilon_{23}({\bf k}) & \epsilon_{24}({\bf k}) \\
\hline
\epsilon_{31}({\bf k}) & \epsilon_{32}({\bf k}) & \epsilon_{33}({\bf k}) & \epsilon_{34}({\bf k}) \\
\epsilon_{41}({\bf k}) & \epsilon_{42}({\bf k}) & \epsilon_{43}({\bf
k}) & \epsilon_{44}({\bf k})
\end{array} \right)
\end{align}
where
\begin{align}
& \epsilon_{11}({\bf k}) = -\mu-4t\left( \alpha({\bf k})
 +\frac{I_{13}({\bf k})}{I_{11}} \right)
\end{align}
\begin{align}
& \epsilon_{12}({\bf k}) = -4t\left( \alpha({\bf k})+\frac{I_{24}({\bf k})}{I_{22}}\right)\\
& \epsilon_{21}({\bf k}) = 4t\frac{I_{13}({\bf k})}{I_{11}}\\
& \epsilon_{22}({\bf k}) = -\mu+U+4t\frac{I_{24}({\bf k})}{I_{22}}\\
& \epsilon_{13}({\bf k}) = \epsilon_{14}({\bf k}) =
-\epsilon_{23}({\bf k}) = -\epsilon_{24}({\bf k}) = -4t \\
& \epsilon_{31}({\bf k}) = -4t\frac{\bar{I}_{33}({\bf k})+\bar{I}_{34}({\bf k})}{I_{11}}\\
& \epsilon_{32}({\bf k}) = 4t\frac{\bar{I}_{33}({\bf k})+\bar{I}_{34}({\bf k})}{I_{22}}\\
& \epsilon_{41}({\bf k}) = -4t\frac{\bar{I}_{43}({\bf
k})+\bar{I}_{44}({\bf
k})}{I_{11}}\\
& \epsilon_{42}({\bf k}) = 4t\frac{\bar{I}_{43}({\bf k})
 +\bar{I}_{44}({\bf k})}{I_{22}}
\end{align}
and
\begin{equation}\label{ebb}
\epsilon_{BB}({\bf k})=m_{BB}({\bf k})I^{-1}_{BB}({\bf k})
\end{equation}
where
\begin{align}\label{mbb}
m_{BB}({\bf k}) & = \mathcal{F}\langle\{ i\frac{\partial}{\partial t}
 \bar{\psi}_{B\sigma}({\bf i},t),
\bar{\psi}^{\dagger}_{B\sigma}({\bf j},t) \}\rangle \nonumber \\
& = \left(
\begin{array}{cc}
m_{33}({\bf k}) & m_{34}({\bf k}) \\
m_{34}({\bf k}) & m_{44}({\bf k})
\end{array} \right).
\end{align}

The operator $\bar{\psi}_{B\sigma}(i)$ provides two-site excitations
and represents the natural extension of $\psi_{A\sigma}(i)$,
which describes one-site excitations,
in the sense of a series expansion over finite-cluster excitations.
Equations of motion of $\bar{\psi}_{B\sigma}(i)$ are much more lengthy
and have a much more complex form with respect to those of
$\psi_{A\sigma}(i)$ as they contain many three-site composite operators.
The application of a systematic
projection/truncation procedure, as the one applied to the equations of
motion of $\psi_{A\sigma}(i)$, is just unfeasible in this case as,
besides to be very lengthy, it would lead to the appearance of a plenty of
unknown correlation functions in the energy matrix. In order to fix all these
correlation functions, we would be forced to use some decoupling
and would completely lose any possibility to control the approximation.
Then, we have opted for a controlled, at least in philosophy,
approximation at the level of equations of motion and decided to neglect
irreducible three-site operators by paying attention to evaluate
exactly all one- and two-site components \cite{Odashima_04}.
This choice has only one obvious drawback: the two-site correlations,
not damped by three-site processes, result quite enhanced.
\begin{align}
& \label{em02a}\mathrm{i} \frac{\partial}{\partial t}\xi_{s\sigma}(i) \simeq
-\mu\xi_{s\sigma}(i)+4t\left\{ \frac{1}{2}\eta_{\sigma}(i)
+\xi^{\alpha}_{s\sigma}(i)+2\eta^{\alpha}_{s\sigma}(i) \right\} \\
& \label{em02b}\mathrm{i} \frac{\partial}{\partial t}\eta_{s\sigma}(i) \simeq
(-\mu+U)\eta_{s\sigma}(i)+4t\left\{\frac{1}{4}\eta_{\sigma}(i)
+\xi^{\alpha}_{s\sigma}(i) \right\}.
\end{align}
\begin{figure}[tbp]
\includegraphics[width=0.43\textwidth]{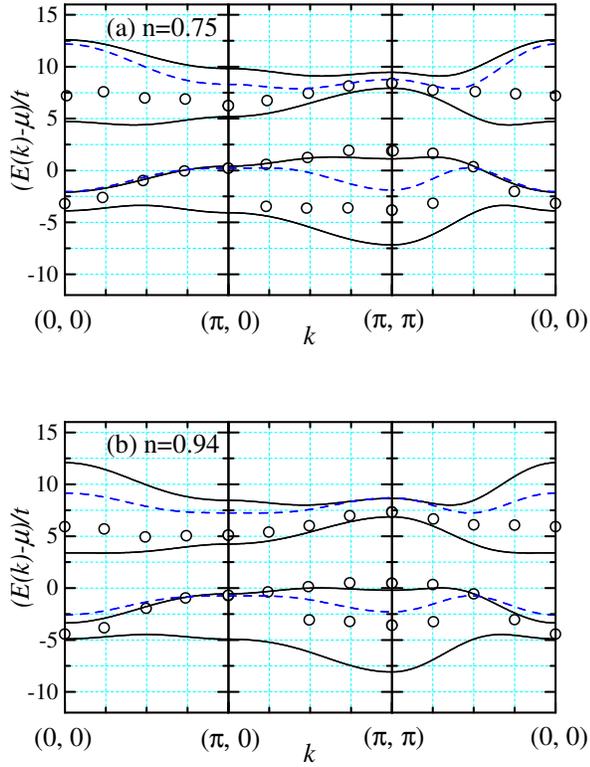}
\caption{The dispersion relation at $U/t=8$, $T/t=0.5$ and $n=0.75$
and $0.94$. The 2-pole solution (dashed line) and QMC data (circle)
of Ref. \onlinecite{Bulut_94a} are also reported.}\label{fig03}
\end{figure}

\begin{figure}[tbp]
\includegraphics[width=0.43\textwidth]{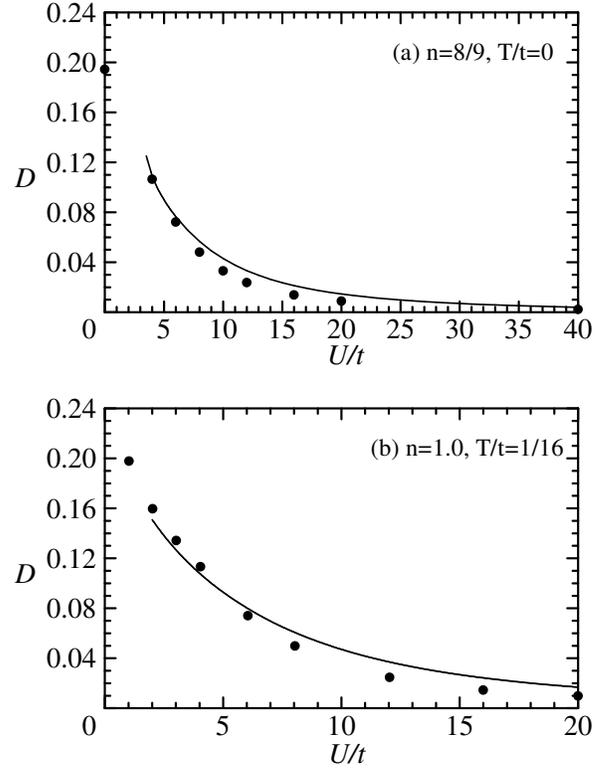}
\caption{Double occupancy (solid line) versus $U/t$ at (a) $n=8/9$
and $T/t=0$ (Lanczos data (circle) from Ref. \onlinecite{Becca_00})
and (b) $n=1.0$ and $T/t=1/16$ (QMC data (circle) from
Ref. \onlinecite{White_89}).}\label{fig04}
\end{figure}
Equations of motion (\ref{em02a}) and (\ref{em02b}) give a simplified form of
$\epsilon_{BB}({\bf k})$
\begin{align}\label{ebb02}
& \epsilon_{33}({\bf k}) = -\mu+4t\left( \alpha({\bf k})+\frac{I_{13}({\bf k})}{I_{11}}\right)\\
& \epsilon_{34}({\bf k}) = 4t\left( 2\alpha({\bf k})+\frac{I_{13}({\bf k})}{I_{11}}\right)\\
& \epsilon_{43}({\bf k}) = 4t\left( \alpha({\bf
k})-\frac{I_{24}({\bf
k})}{I_{22}}\right)\\
& \epsilon_{44}({\bf k}) = -\mu+U-4t\frac{I_{24}({\bf k})}{I_{22}}.
\end{align}
The correlation functions appearing in $I$ and $\epsilon$, except
for $p_{\sigma}$, can be now self-consistently determined through
the retarded Green's function
\begin{equation}
\langle \psi\psi^{\dagger}\rangle=\left( \frac{a}{2\pi}\right)^{2}
\iint d{\bf k} \, d\omega \, \left[1-f_{F}(\omega)\right] \left(
-\frac{1}{\pi}\Im\left[ G(\omega, {\bf k})\right]\right)
\end{equation}
where $f_{F}(\omega)$ is the Fermi distribution function.

The parameters $p_{\sigma}$ are out of the scheme of the present
formulation as they cannot be directly connected to the Green's
function under study. They describe nearest-neighbor spin, charge
and pair correlations and, according to their actual values, play a
fundamental role in determining the behavior of the system
(antiferromagnetic character of the band dispersion, presence of
metal-insulator transition, etc.).\cite{Mancini_04}
The use of operators not satisfying canonical
commutation relations leads to the appearance of unknown correlation functions
in the formulation. The presence of these unknown correlation function
should not be seen as an inconvenience, as many other formulations do,
but as an opportunity given by the method to implement exact relations
dictated by symmetries and/or general principles
which are not automatically satisfied, that is, which are
no more embedded in the Hilbert space of the composite operators
whose Green's functions are computed.\cite{Mancini_00,Mancini_04}
According to this, we will evaluate $p_{\sigma}$ by means of
the algebra constraints \cite{Mancini_00,Mancini_04} $\langle
\xi^{\dagger}_{\sigma}\eta_{\sigma}\rangle =0$. These constraints ensure
that no state referring to a \emph{triple}-occupied site (obviously
forbidden by the Pauli principle) is taken into account in the averaging
procedure and give the possibility to make the correct spin/particle
counting. This allows to fulfil the particle-hole symmetry and
to correctly describe the virtual processes at the basis of
the low-energy processes ($J$ scale of energy).
A comprehensive analysis by means of the two-pole approximation
has shown that the use of algebraic constraints provides
very good agreement with the numerical simulation results well
beyond the conventional Hubbard-I and Roth's decoupling scheme.
\cite{Mancini_04}

\begin{figure}[tbp]
\includegraphics[width=0.43\textwidth]{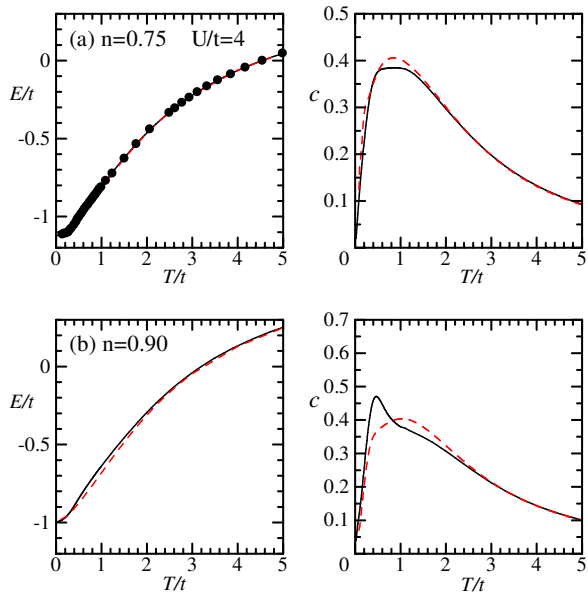}
\caption{Internal energy and specific heat at $U/t=4$ and $n=0.75$
and $0.90$ (solid line). Finite temperature Lanczos data (dashed
line) from Ref.~\onlinecite{Bonca_02} and QMC data (circle) from
Ref. \onlinecite{Duffy_97} are also reported.}\label{fig05}
\end{figure}

\section{BAND STRUCTURE}\label{sec_DOS}

Figure~\ref{fig01} shows the density of states and the corresponding
dispersion relation for $U/t=8$, $n=0.9$ and $T/t=0.01$. As a first
consequence of choosing a basis constituted of four operators, we
have a four-band structure. Together with the usual Hubbard
splitting of the non-interacting band with the appearance of a gap
of the order $U$ between the Hubbard subbands, we can clearly
observe other two dispersion lines in the lower and higher energy
regions ($\omega/t \sim -4$ and $8$). They can be interpreted as
shadow bands coming from the antiferromagnetic nature of the spin
fluctuations. In fact, they show a tendency towards a doubling of
the zone through a mirroring of the original dispersion lines.
Another remarkable feature is the presence of a well developed peak
structure at the Fermi level which comes from the band flatness
around the ($\pi$,$0$)-point. Some peculiar features of the 2-pole
solution (e.g., the inflexion around the ($\pi$,$\pi$)-point) can be
now clearly interpreted as due to the necessity of miming the
behavior of both Hubbard subbands and shadow bands by means of only
two bands.

\begin{figure}[tbp]
\includegraphics[width=0.43\textwidth]{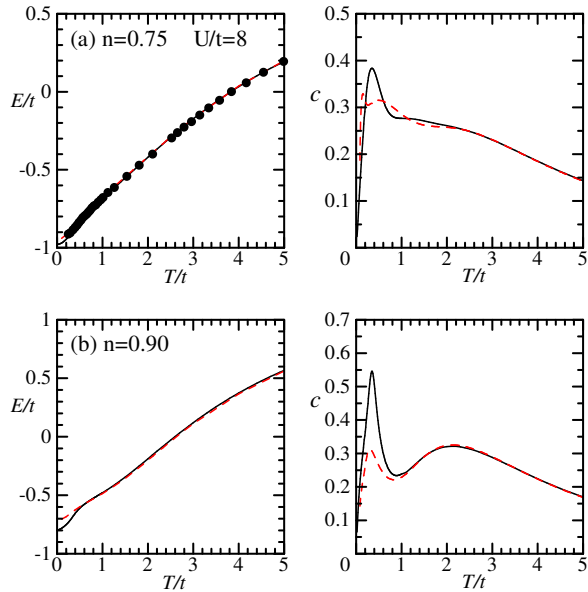}
\caption{Same of Fig. \ref{fig05} but at $U/t=8$.}\label{fig06}
\end{figure}

In Fig.~\ref{fig02}, we present the doping dependence of the density
of states and of the corresponding dispersion relation. The Fermi
level stays pinned to the band flatness located at ($\pi$,$0$)-point
within a wide range of hole-doping ($0 \leq \delta \leq 0.15$). For
higher doping, it moves towards higher energies giving an
electron-like Fermi surface centered at ($0$,$0$)-point. Those
characteristics are commonly observed in many theoretical analysis
\cite{Matsumoto_96a,Matsumoto_97,Krivenko_04,Onoda_03,Dorneich_99,Odashima_04}
which take into account nearest-neighbor spin and charge
fluctuations, in agreement with QMC
data.\cite{Bulut_94a,Preuss_95,Preuss_97} Therefore, we can conclude
that in order to reproduce such peculiar features, it is necessary
to take into account high-order operators in the basis, as we have
done in this manuscript.

In Fig.~\ref{fig03}, we provide a detailed comparison of the band
structure obtained by the present formulation with QMC results. As
can be easily seen, we have a good agreement with QMC data,
especially for the low-energy band around the Fermi level. We
observe shadow bands more pronounced than in QMC data. We should
recall that the present formulation is a pole-approximation and
damping effects are neglected. Furthermore, three-site terms in the
equations of motion have been neglected. Therefore, there is no
diffusion process to weaken two-site correlation effects. On the
other hand, we can expect that with hole-doping the
antiferromagnetic correlations are weakened by hole motion and that
the shadow bands become broader and broader owing to damping
effects. Eventually, shadow bands are wiped out and we may observe
traces of them as shoulders of the main Hubbard bands, as seen in
QMC results.

\begin{figure}[tbp]
\includegraphics[width=0.38\textwidth]{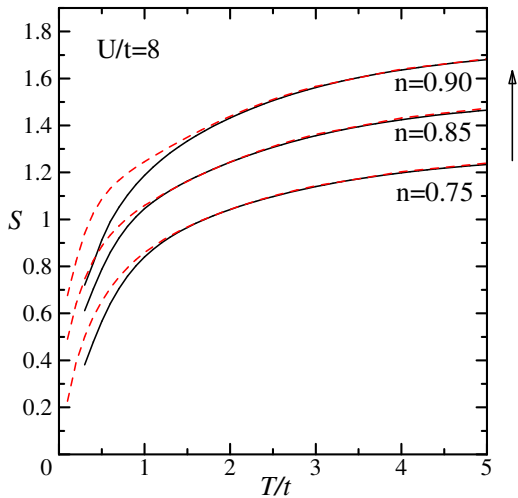}
\caption{Entropy at $U/t=8$ (solid line) (Finite temperature Lanczos
data (dashed line) from Ref. \onlinecite{Bonca_02}). Data are
shifted of $0.2$ along the vertical axis for the sake of
clarity.}\label{fig07}
\end{figure}

\section{THERMODYNAMIC QUANTITIES}\label{sec_THERMODYNAMICS}

Figure~\ref{fig04} presents double occupancy $D=\langle
n_{\uparrow}n_{\downarrow}\rangle=\langle
\eta^{\dagger}_{\sigma}\eta_{\sigma}\rangle$ in comparison with
Lanczos \cite{Becca_00} and QMC \cite{White_89} results. For
$U/t<3$, it is difficult to get self-consistent solution as split
bands start merging. Our results show a good agreement in a wide
range of $U/t$ values. We should point out that the 2-pole
approximation also provides similar agreement.\cite{Mancini_04}

Internal energy $E=\langle H\rangle/N$ and specific heat $C=dE/dT$
per site at $U/t=4$ and $8$ and $n=0.75$ and $0.90$ are reported in
Figs.~\ref{fig05} and \ref{fig06}. Data from finite temperature
Lanczos \cite{Bonca_02} and QMC \cite{Duffy_97} are provided for
comparison. It is worth mentioning that our results for $U/t=12$
have the same general features that those for $U/t=8$, but with more
pronounced characteristics. As regards the internal energy, the
agreement with the Lanczos data is excellent except for the low
temperature region $T/t < 0.4$ at $U/t \ge 8$. As regards the
specific heat, we observe a sharp peak around $T/t \sim 0.3$ and a
fairly broad peak in the higher temperature region $T/t>1$. The two
peak structure is more pronounced for $U/t \ge 8$, but not so much
for $U/t=4$. This tendency is also observed in several numerical
simulations.\cite{Bonca_02,Duffy_97,Paiva_01} Usually, the sharp
peak at lower temperatures and the broad peak at higher temperatures
are interpreted as consequences of spin and charge fluctuations
related to the energy scales of $J$ and $U$, respectively. The main
difference between our results and numerical ones regards the height
of the peak in the specific heat around $T/t \sim 0.3$ that comes
from the decrease in the internal energy. This is an indication of
well established spin ordering which cannot be correctly evaluated
on a small cluster. Numerical simulation cannot describe spin and
charge ordering in the case that the correlation lengths exceed the
cluster size. On the other hand, in our formulation, as already
discussed in Sec.~\ref{sec_DOS}, there is no diffusion process to
weaken two-site correlation effects. Therefore, there is a tendency
to have too pronounced spin and charge correlations.

\begin{figure}[tbp]
\includegraphics[width=0.40\textwidth]{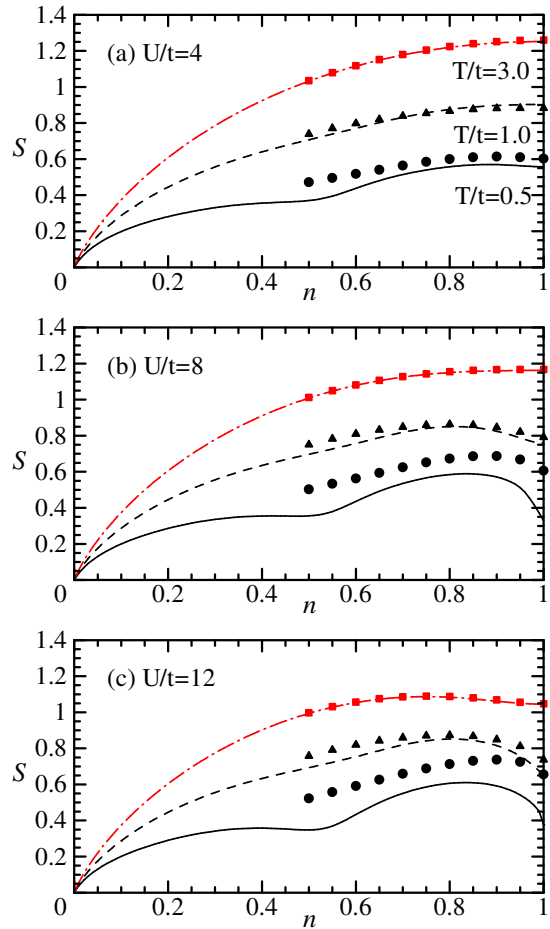}
\caption{Entropy vs $n$ at $U/t=4$, $8$ and $12$ and $T/t=0.5$,
$1.0$ and $3.0$ (lines) (Finite temperature Lanczos data (markers)
from Ref. \onlinecite{Bonca_02}).}\label{fig08}
\end{figure}

We can discuss this issue in more detail by commenting our results
on entropy
\begin{equation}\label{s02}
S(T, n)=-\int^{n}_{0}dn^{\prime}\left( \frac{\partial \mu}{\partial
T}\right)_{n^{\prime}}.
\end{equation}
This relation is derived from the thermodynamic relations
\cite{Mancini_99a,Mancini_04} $S=-\left( \frac{\partial F}{\partial
T}\right)_{n}$ and $\mu=\left( \frac{\partial F}{\partial
n}\right)_{T}$, which give the Maxwell relation $\left(
\frac{\partial S}{\partial n}\right)_{T} =-\left( \frac{\partial
\mu}{\partial T}\right)_{n}$.
Figure~\ref{fig07} reports the results of entropy in comparison with
Lanczos ones. We present results only for $U/t=8$, but $U/t=4$ and
$12$ results show the same tendency. Our results completely coincide
with Lanczos ones in the high-temperature region ($T/t>1$). In the
low-temperature region ($T/t<1$), our data are lower than Lanczos
ones indicating a stronger tendency to ordering.

To better understand the tendency to ordering, we have investigated
the filling dependence of entropy at several temperatures (see
Fig.~\ref{fig08}). For $T/t \ge 1.0$, the agreement is extremely
good in whole range of filling. However, at lower temperatures, our
results show a decrease around both quarter and half filling.
Usually, a decrease of entropy at quarter and half filling is
interpreted as an indication of charge and spin ordering,
respectively. In a small cluster, it is difficult to investigate
such ordered states because of the system size.

\section{SUMMARY}\label{sec_SUMMARY}

In the present paper, we have carried out an analysis of the
two-dimensional Hubbard model by means of a 4-pole approximation
within the Composite Operator Method. Density of states and
corresponding dispersion relation show remarkable characteristics:
four-bands structure, a quasi-particle peak at the Fermi level,
shadow bands and band flatness at ($\pi$, $0$)-point. Quantities
such as double occupancy, internal energy, specific heat and entropy
have been comprehensively investigated. Our results show an
excellent agreement with numerical simulations except for the low
energy features related to spin and charge ordering. In the present
formulation, nearest-neighbor-site effects are probably
over-estimated whereas on a small cluster there are some
difficulties to describe spin and charge ordering in the case that
the correlation length exceeds the system size. Probably the right
stays in the middle: our results provide meaningful information for
band structure and thermodynamic quantities, but the inclusion of
damping effects is necessary for complete understanding of this
system.

\appendix

\begin{widetext}

\appendix

\section{}

The equations of motion of $\xi_{s\sigma}(i)$ and
$\eta_{s\sigma}(i)$ reads as
\begin{equation}\label{emxis}
\begin{array}{l}
\displaystyle \mathrm{i}\frac{\partial}{\partial t} \xi_{s\sigma}(i) =-\mu \xi_{s\sigma}(i) \\
\displaystyle \hspace*{5mm} -4t\left\{ \begin{array}{l}
 -n_{-\sigma}(i)c^{\alpha^{2}}_{\sigma}(i)
 +n_{-\sigma}(i)(n_{-\sigma}(i)c^{\alpha}_{\sigma}(i))^{\alpha}
 -n_{-\sigma}(i)(c^{\dagger}_{-\sigma}(i)c_{\sigma}(i)c^{\alpha}_{-\sigma}(i))^{\alpha} \\
 +n_{-\sigma}(i)(c^{\alpha\dagger}_{-\sigma}(i)c_{\sigma}(i)c_{-\sigma}(i))^{\alpha}
 -c^{\dagger}_{-\sigma}(i)c^{\alpha}_{-\sigma}(i)\xi^{\alpha}_{\sigma}(i)
 +c^{\alpha\dagger}_{-\sigma}(i)c_{-\sigma}(i)\xi^{\alpha}_{\sigma}(i) \\
 +c^{\dagger}_{-\sigma}(i)c_{\sigma}(i)c^{\alpha^{2}}_{-\sigma}(i)
 -c^{\dagger}_{-\sigma}(i)c_{\sigma}(i)(n_{\sigma}(i)c^{\alpha}_{-\sigma}(i))^{\alpha}
 +c^{\dagger}_{-\sigma}(i)c_{\sigma}(i)(c^{\dagger}_{\sigma}(i)c_{-\sigma}(i)c^{\alpha}_{\sigma}(i))^{\alpha} \\
 -c^{\dagger}_{-\sigma}(i)c_{\sigma}(i)(c^{\alpha\dagger}_{\sigma}(i)c_{-\sigma}(i)c_{\sigma}(i))^{\alpha}
 +c^{\dagger}_{-\sigma}(i)c^{\alpha}_{\sigma}(i)\xi^{\alpha}_{-\sigma}(i)
 -c^{\alpha\dagger}_{-\sigma}(i)c_{\sigma}(i)\xi^{\alpha}_{-\sigma}(i) \\
 -\eta^{\alpha\dagger}_{-\sigma}(i)c_{\sigma}(i)c^{\alpha}_{-\sigma}(i)
 -\eta^{\alpha\dagger}_{-\sigma}(i)c^{\alpha}_{\sigma}(i)c_{-\sigma}(i)
 +(c^{\dagger}_{\sigma}(i)c^{\dagger}_{-\sigma}(i)c^{\alpha}_{\sigma}(i))^{\alpha}c_{\sigma}(i)c_{-\sigma}(i) \\
 +(c^{\alpha\dagger}_{-\sigma}(i)n_{\sigma}(i))^{\alpha}c_{\sigma}(i)c_{-\sigma}(i)
 +(c^{\alpha\dagger}_{\sigma}(i)c_{\sigma}(i)c^{\dagger}_{-\sigma}(i))^{\alpha}c_{\sigma}(i)c_{-\sigma}(i)
 \end{array} \right\}
\end{array}
\end{equation}
\begin{equation}\label{emetas}
\begin{array}{l}
\displaystyle \mathrm{i}\frac{\partial}{\partial t} \eta_{s\sigma}(i) =
(-\mu+U) \eta_{s\sigma}(i) \\
\displaystyle \hspace*{5mm} -4t\left\{ \begin{array}{l}
 -n_{-\sigma}(i)(n_{-\sigma}(i)c^{\alpha}_{\sigma}(i))^{\alpha}
 +n_{-\sigma}(i)(c^{\dagger}_{-\sigma}(i)c_{\sigma}(i)c^{\alpha}_{-\sigma}(i))^{\alpha}
 -n_{-\sigma}(i)(c^{\alpha\dagger}_{-\sigma}(i)c_{\sigma}(i)c_{-\sigma}(i))^{\alpha} \\
 -c^{\dagger}_{-\sigma}(i)c^{\alpha}_{-\sigma}(i)\eta^{\alpha}_{\sigma}(i)
 +c^{\alpha\dagger}_{-\sigma}(i)c_{-\sigma}(i)\eta^{\alpha}_{\sigma}(i)
 +c^{\dagger}_{-\sigma}(i)c_{\sigma}(i)(n_{\sigma}(i)c^{\alpha}_{-\sigma}(i))^{\alpha} \\
 -c^{\dagger}_{-\sigma}(i)c_{\sigma}(i)(c^{\dagger}_{\sigma}(i)c_{-\sigma}(i)c^{\alpha}_{\sigma}(i))^{\alpha}
 +c^{\dagger}_{-\sigma}(i)c_{\sigma}(i)(c^{\alpha\dagger}_{\sigma}(i)c_{-\sigma}(i)c_{\sigma}(i))^{\alpha} \\
 +c^{\dagger}_{-\sigma}(i)c^{\alpha}_{\sigma}(i)\eta^{\alpha}_{-\sigma}(i)
 -c^{\alpha\dagger}_{-\sigma}(i)c_{\sigma}(i)\eta^{\alpha}_{-\sigma}(i)
 -\xi^{\alpha\dagger}_{-\sigma}(i)c_{\sigma}(i)c^{\alpha}_{-\sigma}(i)
 -\xi^{\alpha\dagger}_{-\sigma}(i)c^{\alpha}_{\sigma}(i)c_{-\sigma}(i) \\
 -(c^{\dagger}_{\sigma}(i)c^{\dagger}_{-\sigma}(i)c^{\alpha}_{\sigma}(i))^{\alpha}c_{\sigma}(i)c_{-\sigma}(i)
 +c^{\alpha^{2}\dagger}_{-\sigma}(i)c_{\sigma}(i)c_{-\sigma}(i)
 -(c^{\alpha\dagger}_{-\sigma}(i)n_{\sigma}(i))^{\alpha}c_{\sigma}(i)c_{-\sigma}(i) \\
 -(c^{\alpha\dagger}_{\sigma}(i)c_{\sigma}(i)c^{\dagger}_{-\sigma}(i))^{\alpha}c_{\sigma}(i)c_{-\sigma}(i)
 \end{array} \right\}
\end{array}
\end{equation}
where $A^{\alpha^{2}}(i)$ and $A^{\alpha}(i)B^{\alpha}(i)$ stand for
\begin{equation}
A^{\alpha^{2}}(i)=\sum_{\bf j,k}\alpha_{\bf ij}\alpha_{\bf jk}A(k)
\end{equation}
and
\begin{equation}
A^{\alpha}(i)B^{\alpha}(i)=\sum_{\bf j,k}\alpha_{\bf
ij}A(j)\alpha_{\bf ik}B(k)
\end{equation}
respectively. We can isolate the single-site terms as follows
\begin{equation}
A^{\alpha^{2}}(i)=\frac{1}{4}A(i)+\sum_{\bf j\neq k}\alpha_{\bf
ij}\alpha_{\bf jk}A(k),
\end{equation}
\begin{equation}
A^{\alpha}(i)B^{\alpha}(i)=\frac{1}{4}\left(A(i)B(i)
\right)^{\alpha}
 +\sum_{\bf j\neq k}\alpha_{\bf ij}A(j)\alpha_{\bf ik}B(k).
\end{equation}

Then, we can isolate the one- and two-site terms in the equations of
motion and neglect high-order terms for simplicity
\begin{equation}
\begin{array}{l}
\displaystyle \mathrm{i}\frac{\partial}{\partial t}\xi_{s\sigma}(i) \simeq
 -\mu\xi_{s\sigma}(i) +4t \left\{ \frac{1}{2}\eta_{\sigma}(i)
 +\xi^{\alpha}_{s\sigma}(i)
 +2\eta^{\alpha}_{s\sigma}(i) \right\} \\
\displaystyle \mathrm{i}\frac{\partial}{\partial t}\eta_{s\sigma}(i) \simeq
 (-\mu+U)\eta_{s\sigma}(i) +4t \left\{
 \frac{1}{4}\eta_{\sigma}(i)+\xi^{\alpha}_{s\sigma}(i) \right\}
\end{array}.
\end{equation}
Recalling the relations between $\bar{\psi}_{\sigma}(i)$ and
$\psi_{\sigma}(i)$
\begin{equation}
\begin{array}{l}
\bar{\xi}_{s\sigma}(i)=\xi_{s\sigma}(i)-A_{31}(-i\nabla)\xi_{\sigma}(i) \\
\bar{\eta}_{s\sigma}(i)=\eta_{s\sigma}(i)-A_{42}(-i\nabla)\eta_{\sigma}(i),
\end{array}
\end{equation}
we have
\begin{equation}
\begin{array}{l}
\displaystyle \mathrm{i}\frac{\partial}{\partial t}\bar{\xi}_{s\sigma}(i)
\simeq
 -\mu \bar{\xi}_{s\sigma}(i) \\
 \displaystyle \hspace*{5mm} -4t\left\{ \begin{array}{l}
 \displaystyle -\left( A_{31}( -i\nabla )\alpha ( -i\nabla )+A_{31}( -i\nabla )A_{31}( -i\nabla )
  +\alpha ( -i\nabla )A_{31}( -i\nabla ) \right) \xi_{\sigma}(i) \\
 \displaystyle -\left( \frac{1}{2}+A_{31}( -i\nabla )\alpha ( -i\nabla )+A_{31}( -i\nabla )A_{42}( -i\nabla )
  +2\alpha ( -i\nabla )A_{42}( -i\nabla ) \right) \eta_{\sigma}(i) \\
 \displaystyle -\left( \alpha ( -i\nabla )+A_{31}( -i\nabla ) \right) \bar{\xi}_{s\sigma}(i)
 -\left( 2\alpha ( -i\nabla )+A_{31}( -i\nabla ) \right) \bar{\eta}_{s\sigma}(i) \end{array} \right\} \\
\displaystyle \mathrm{i}\frac{\partial}{\partial t}\bar{\eta}_{s\sigma}(i)
\simeq
 (-\mu +U) \bar{\eta}_{s\sigma}(i) \\
 \displaystyle \hspace*{5mm} -4t\left\{ \begin{array}{l}
 \displaystyle -\left( \alpha ( -i\nabla )-A_{42}( -i\nabla ) \right)
  A_{31}( -i\nabla ) \xi_{\sigma}(i)
  -\left( \frac{1}{4} -A_{42}( -i\nabla )A_{42}( -i\nabla ) \right) \eta_{\sigma}(i) \\
 \displaystyle -\left( \alpha ( -i\nabla )-A_{42}( -i\nabla ) \right) \bar{\xi}_{s\sigma}(i)
 +A_{42}( -i\nabla )\bar{\eta}_{s\sigma}(i) \end{array} \right\}.
\end{array}
\end{equation}
Then, $\epsilon_{BB}({\bf k})$ can be obtained by simple inspection
\begin{equation}
\epsilon_{BB}({\bf k}) \simeq \left( \begin{array}{cc}
-\mu+4t(\alpha({\bf k})+A_{31}({\bf k})) & 4t(2\alpha({\bf k})+A_{31}({\bf k})) \\
4t(\alpha({\bf k})-A_{42}({\bf k})) & -\mu+U-4tA_{42}({\bf k})
\end{array} \right).
\end{equation}
where $A_{13}$ and $A_{24}$ are defined in the main text.

\section{}

$I_{33}({\bf i,j})$, $I_{34}({\bf i,j})$, and $I_{44}({\bf i,j})$
reads as
\begin{equation}
\begin{array}{l}
I_{33\sigma}({\bf i,j})=\langle \{ \xi_{s\sigma}({\bf i},t),\ \xi^{\dagger}_{s\sigma}({\bf j},t) \} \rangle \\
\displaystyle \hspace*{5mm} =\delta_{\bf ij} \left( \begin{array}{l}
 \langle \eta^{\alpha\dagger}_{-\sigma}(i)\eta^{\alpha}_{-\sigma}(i) \rangle
 +\langle \xi^{\alpha\dagger}_{-\sigma}(i)\xi^{\alpha}_{-\sigma}(i)n_{\sigma}(i) \rangle
 -\langle \xi^{\alpha\dagger}_{-\sigma}(i)\xi^{\alpha}_{-\sigma}(i)n_{-\sigma}(i) \rangle
 -\langle \eta^{\alpha\dagger}_{-\sigma}(i)\eta^{\alpha}_{-\sigma}(i)n_{\sigma}(i) \rangle \\
 -\langle \eta^{\alpha\dagger}_{-\sigma}(i)\eta^{\alpha}_{-\sigma}(i)n_{-\sigma}(i) \rangle
 -\langle \xi^{\alpha\dagger}_{\sigma}(i)\eta^{\alpha\dagger}_{-\sigma}(i)c_{\sigma}(i)c_{-\sigma}(i) \rangle
 +\langle c^{\dagger}_{-\sigma}(i)c^{\dagger}_{\sigma}(i)
  \xi^{\alpha}_{\sigma}(i)\eta^{\alpha}_{-\sigma}(i) \rangle \\
 -\langle \xi^{\alpha\dagger}_{\sigma}(i)\xi^{\alpha}_{-\sigma}(i)c^{\dagger}_{-\sigma}(i)c_{\sigma}(i) \rangle
 -\langle \xi^{\alpha\dagger}_{-\sigma}(i)\xi^{\alpha}_{\sigma}(i)c^{\dagger}_{\sigma}(i)c_{-\sigma}(i) \rangle
 \end{array} \right) \\
\displaystyle \hspace*{5mm} +\alpha_{\bf ij} \left( \begin{array}{l}
 \langle \xi^{\dagger}_{-\sigma}(i)\xi^{\alpha}_{-\sigma}(i)n_{-\sigma}(j) \rangle
 -\langle \eta^{\alpha\dagger}_{-\sigma}(i)\eta_{-\sigma}(i)n_{-\sigma}(j) \rangle
 +\langle \xi^{\alpha\dagger}_{-\sigma}(j)\xi_{-\sigma}(j)n_{-\sigma}(i) \rangle \\
 -\langle n_{-\sigma}(i)\eta^{\dagger}_{-\sigma}(j)\eta^{\alpha}_{-\sigma}(j) \rangle
 +\langle \xi^{\dagger}_{-\sigma}(i)\xi^{\alpha}_{\sigma}(i)c^{\dagger}_{\sigma}(j)c_{-\sigma}(j) \rangle
 -\langle \eta^{\alpha\dagger}_{-\sigma}(i)\eta_{\sigma}(i)c^{\dagger}_{\sigma}(j)c_{-\sigma}(j) \rangle \\
 +\langle \xi^{\alpha\dagger}_{\sigma}(j)\xi_{-\sigma}(j)c^{\dagger}_{-\sigma}(i)c_{\sigma}(i) \rangle
 -\langle c^{\dagger}_{-\sigma}(i)c_{\sigma}(i)\eta^{\dagger}_{\sigma}(j)
  \eta^{\alpha}_{-\sigma}(j) \rangle
 +\langle \xi^{\alpha\dagger}_{\sigma}(j)\eta^{\dagger}_{-\sigma}(j)c_{\sigma}(i)c_{-\sigma}(i) \rangle \\
 -\langle \xi^{\alpha\dagger}_{-\sigma}(j)\eta^{\dagger}_{\sigma}(j)c_{\sigma}(i)c_{-\sigma}(i) \rangle
 +\langle c^{\dagger}_{-\sigma}(j)c^{\dagger}_{\sigma}(j)\eta_{-\sigma}(i)\xi^{\alpha}_{\sigma}(i) \rangle
 -\langle c^{\dagger}_{-\sigma}(j)c^{\dagger}_{\sigma}(j)\eta_{\sigma}(i)\xi^{\alpha}_{-\sigma}(i) \rangle
 \end{array} \right) \\
\displaystyle \hspace*{5mm} +\sum_{\bf k}\alpha_{\bf ik}\alpha_{\bf
kj} \left( \begin{array}{l}
  \langle n_{-\sigma}(i)n_{-\sigma}(j) \rangle
 -\langle n_{-\sigma}(i)n_{-\sigma}(k)n_{-\sigma}(j) \rangle
 -\langle c^{\dagger}_{-\sigma}(i)c_{\sigma}(i)c^{\dagger}_{\sigma}(k)c_{-\sigma}(k)n_{-\sigma}(j) \rangle \\
 +\langle c_{\sigma}(i)c_{-\sigma}(i)c^{\dagger}_{-\sigma}(k)c^{\dagger}_{\sigma}(k)n_{-\sigma}(j) \rangle
 -\langle n_{-\sigma}(i)c^{\dagger}_{-\sigma}(k)c_{\sigma}(k)c^{\dagger}_{\sigma}(j)c_{-\sigma}(j) \rangle \\
 +\langle c^{\dagger}_{-\sigma}(i)c_{\sigma}(i)c^{\dagger}_{j\sigma}(j)c_{-\sigma}(j) \rangle
 -\langle c^{\dagger}_{-\sigma}(i)c_{\sigma}(i)n_{\sigma}(k)c^{\dagger}_{\sigma}(j)c_{-\sigma}(j) \rangle \\
 +\langle c^{\dagger}_{-\sigma}(j)c^{\dagger}_{\sigma}(j)c_{\sigma}(k)c_{-\sigma}(k)n_{-\sigma}(i) \rangle
 +\langle c^{\dagger}_{-\sigma}(j)c^{\dagger}_{\sigma}(j)n_{\sigma}(k)c_{\sigma}(i)c_{-\sigma}(i) \rangle
 \end{array} \right)
\end{array}
\end{equation}
\begin{equation}
\begin{array}{l}
I_{34\sigma}({\bf i,j})=\langle \{ \xi_{s\sigma}({\bf i},t),\ \eta^{\dagger}_{s\sigma}({\bf j},t) \} \rangle \\
\hspace*{5mm} =\delta_{\bf ij} \left( \begin{array}{l}
 \langle \eta^{\alpha\dagger}_{-\sigma}(i)\xi^{\alpha}_{-\sigma}(i) \rangle
 -2\langle \eta^{\alpha\dagger}_{-\sigma}(i)\xi^{\alpha}_{-\sigma}(i)n_{-\sigma}(i) \rangle
 -\langle \eta^{\alpha\dagger}_{\sigma}(i)\xi^{\alpha}_{-\sigma}(i)c^{\dagger}_{-\sigma}(i)c_{\sigma}(i) \rangle \\
 -\langle \eta^{\alpha\dagger}_{-\sigma}(i)\xi^{\alpha}_{\sigma}(i)c^{\dagger}_{\sigma}(i)c_{-\sigma}(i) \rangle
 +\langle \eta^{\alpha\dagger}_{\sigma}(i)\eta^{\alpha\dagger}_{-\sigma}(i)c_{-\sigma}(i)c_{\sigma}(i) \rangle
 +\langle c^{\dagger}_{-\sigma}(i)c^{\dagger}_{\sigma}(i)\xi^{\alpha}_{\sigma}(i)\xi^{\alpha}_{-\sigma}(i) \rangle
 \end{array} \right) \\
\hspace*{5mm} +\alpha_{\bf ij} \left( \begin{array}{l}
 -\langle \eta^{\alpha\dagger}_{-\sigma}(i)\xi_{-\sigma}(i)n_{-\sigma}(j) \rangle
 +\langle \eta^{\dagger}_{-\sigma}(i)\xi^{\alpha}_{-\sigma}(i)n_{-\sigma}(j) \rangle
 +\langle \eta^{\alpha\dagger}_{-\sigma}(j)\xi_{-\sigma}(j)n_{-\sigma}(i) \rangle \\
 -\langle n_{-\sigma}(i)\eta^{\dagger}_{-\sigma}(j)\xi^{\alpha}_{-\sigma}(j) \rangle
 +\langle \eta^{\dagger}_{-\sigma}(i)\xi^{\alpha}_{\sigma}(i)c^{\dagger}_{\sigma}(j)c_{-\sigma}(j) \rangle
 -\langle \eta^{\alpha\dagger}_{-\sigma}(i)\xi_{\sigma}(i)c^{\dagger}_{\sigma}(j)c_{-\sigma}(j) \rangle \\
 +\langle \eta^{\alpha\dagger}_{\sigma}(j)\xi_{-\sigma}(j)c^{\dagger}_{-\sigma}(i)c_{\sigma}(i) \rangle
 -\langle c^{\dagger}_{-\sigma}(i)c_{\sigma}(i)\eta^{\dagger}_{\sigma}(j)\xi^{\alpha}_{-\sigma}(j) \rangle
 +\langle \eta^{\alpha\dagger}_{\sigma}(j)\eta^{\dagger}_{-\sigma}(j)c_{\sigma}(i)c_{-\sigma}(i) \rangle \\
 -\langle \eta^{\alpha\dagger}_{-\sigma}(j)\eta^{\dagger}_{\sigma}(j)c_{\sigma}(i)c_{-\sigma}(i) \rangle
 -\langle c^{\dagger}_{-\sigma}(j)c^{\dagger}_{\sigma}(j)\xi_{\sigma}(i)\xi^{\alpha}_{-\sigma}(i) \rangle
 +\langle c^{\dagger}_{-\sigma}(j)c^{\dagger}_{\sigma}(j)\xi_{-\sigma}(i)\xi^{\alpha}_{\sigma}(i) \rangle
 \end{array} \right)
\end{array}
\end{equation}
\begin{equation}
\begin{array}{l}
I_{44\sigma}({\bf i,j})=\langle \{ \eta_{s\sigma}({\bf i},t),\ \eta^{\dagger}_{s\sigma}({\bf j},t) \} \rangle \\
\displaystyle \hspace*{5mm} =\delta_{\bf ij} \left( \begin{array}{l}
 \langle \xi^{\alpha\dagger}_{-\sigma}(i)\xi^{\alpha}_{-\sigma}(i) \rangle
 -\langle \xi^{\alpha\dagger}_{-\sigma}(i)\xi^{\alpha}_{-\sigma}(i)n_{\sigma}(i) \rangle
 -\langle \xi^{\alpha\dagger}_{-\sigma}(i)\xi^{\alpha}_{-\sigma}(i)n_{-\sigma}(i) \rangle \\
 +\langle \eta^{\alpha\dagger}_{-\sigma}(i)\eta^{\alpha}_{-\sigma}(i)n_{\sigma}(i) \rangle
 -\langle \eta^{\alpha\dagger}_{-\sigma}(i)\eta^{\alpha}_{-\sigma}(i)n_{-\sigma}(i) \rangle
 -\langle \eta^{\alpha\dagger}_{\sigma}(i)\eta^{\alpha}_{-\sigma}(i)c^{\dagger}_{-\sigma}(i)c_{\sigma}(i) \rangle \\
 -\langle \eta^{\alpha\dagger}_{-\sigma}(i)\eta^{\alpha}_{\sigma}(i)c^{\dagger}_{\sigma}(i)c_{-\sigma}(i) \rangle
 -\langle \eta^{\alpha\dagger}_{\sigma}(i)\xi^{\alpha\dagger}_{-\sigma}(i)c_{\sigma}(i)c_{-\sigma}(i) \rangle
 +\langle c^{\dagger}_{-\sigma}(i)c^{\dagger}_{\sigma}(i)\eta^{\alpha}_{\sigma}(i)\xi^{\alpha}_{-\sigma}(i) \rangle
 \end{array} \right) \\
\displaystyle \hspace*{5mm} +\alpha_{\bf ij} \left( \begin{array}{l}
 -\langle \xi^{\alpha\dagger}_{-\sigma}(i)\xi_{-\sigma}(i)n_{-\sigma}(j) \rangle
 +\langle \eta^{\dagger}_{-\sigma}(i)\eta^{\alpha}_{-\sigma}(i) n_{-\sigma}(j) \rangle
 +\langle \eta^{\alpha\dagger}_{-\sigma}(j)\eta_{-\sigma}(j)n_{-\sigma}(i) \rangle \\
 -\langle n_{-\sigma}(i)\xi^{\dagger}_{-\sigma}(j)\xi^{\alpha}_{-\sigma}(j) \rangle
 -\langle \xi^{\alpha\dagger}_{-\sigma}(i)\xi_{\sigma}(i)c^{\dagger}_{\sigma}(j)c_{-\sigma}(j) \rangle
 +\langle \eta^{\dagger}_{-\sigma}(i)\eta^{\alpha}_{\sigma}(i)c^{\dagger}_{\sigma}(j)c_{-\sigma}(j) \rangle \\
 +\langle \eta^{\alpha\dagger}_{\sigma}(j)\eta_{-\sigma}(j)c^{\dagger}_{-\sigma}(i)c_{\sigma}(i) \rangle
 -\langle c^{\dagger}_{-\sigma}(i)c_{\sigma}(i)\xi^{\dagger}_{\sigma}(j)\xi^{\alpha}_{-\sigma}(j) \rangle
 +\langle c^{\dagger}_{-\sigma}(j)c^{\dagger}_{\sigma}(j)\xi_{-\sigma}(i)\eta^{\alpha}_{\sigma}(i) \rangle \\
 -\langle c^{\dagger}_{-\sigma}(j)c^{\dagger}_{\sigma}(j)\xi_{\sigma}(i)\eta^{\alpha}_{-\sigma}(i) \rangle
 +\langle \eta^{\alpha\dagger}_{\sigma}(j)\xi^{\dagger}_{-\sigma}(j)c_{\sigma}(i)c_{-\sigma}(i) \rangle
 -\langle \eta^{\alpha\dagger}_{-\sigma}(j)\xi^{\dagger}_{\sigma}(j)c_{\sigma}(i)c_{-\sigma}(i) \rangle
 \end{array} \right) \\
\displaystyle \hspace*{5mm} +\sum_{\bf k}\alpha_{\bf ik}\alpha_{\bf
kj} \left( \begin{array}{l}
  \langle n_{-\sigma}(i)n_{-\sigma}(k)n_{-\sigma}(j) \rangle
 +\langle c^{\dagger}_{-\sigma}(i)c_{\sigma}(i)c^{\dagger}_{\sigma}(k)c_{-\sigma}(k)n_{-\sigma}(j) \rangle \\
 +\langle c^{\dagger}_{-\sigma}(i)c_{\sigma}(i)n_{\sigma}(k)c^{\dagger}_{\sigma}(j)c_{-\sigma}(j) \rangle
 +\langle n_{-\sigma}(i)c^{\dagger}_{-\sigma}(k)c_{\sigma}(k)c^{\dagger}_{\sigma}(j)c_{-\sigma}(j) \rangle \\
 -\langle c_{\sigma}(i)c_{-\sigma}(i)c^{\dagger}_{-\sigma}(k)c^{\dagger}_{\sigma}(k)n_{-\sigma}(j) \rangle
 +\langle c^{\dagger}_{-\sigma}(j)c^{\dagger}_{\sigma}(j)c_{-\sigma}(k)c_{\sigma}(k)n_{-\sigma}(i) \rangle \\
 +\langle c^{\dagger}_{-\sigma}(j)c^{\dagger}_{\sigma}(j)c_{\sigma}(i)c_{-\sigma}(i) \rangle
 -\langle c^{\dagger}_{-\sigma}(j)c^{\dagger}_{\sigma}(j)n_{\sigma}(k)c_{\sigma}(i)c_{-\sigma}(i) \rangle
\end{array} \right)
\end{array}.
\end{equation}
They contain three-site correlation functions which cannot be
directly evaluated in terms of the propagators under analysis.
Therefore, we have decided to decouple them in terms of two-site
correlation functions. For example,
\begin{equation}
\begin{array}{l}
\langle c^{\dagger}_{-\sigma}(i)c^{\dagger}_{\sigma}(i)
 \xi^{\alpha}_{\sigma}(i)\eta^{\alpha}_{-\sigma}(i)\rangle \\
\displaystyle \hspace*{5mm}=\frac{1}{4}\langle
c^{\dagger}_{-\sigma}(i)c^{\dagger}_{\sigma}(i)
 \left( \xi_{\sigma}(i)\eta_{-\sigma}(i)\right)^{\alpha}\rangle
 +\sum_{\bf j\neq k}\alpha_{\bf ij}\alpha_{\bf ik}\langle c^{\dagger}_{-\sigma}(i)c^{\dagger}_{\sigma}(i)
 \xi_{\sigma}(j)\eta_{-\sigma}(k)\rangle \\
\displaystyle \hspace*{5mm}\simeq \frac{1}{4}\langle
c^{\dagger}_{-\sigma}(i)c^{\dagger}_{\sigma}(i)
 \left( \xi_{\sigma}(i)\eta_{-\sigma}(i)\right)^{\alpha}\rangle
 +\sum_{\bf j\neq k}\alpha_{\bf ij}\alpha_{\bf ik}\langle c^{\dagger}_{-\sigma}(i)\eta_{-\sigma}(k)\rangle
 \langle c^{\dagger}_{\sigma}(i)\xi_{\sigma}(j)\rangle.
\end{array}
\end{equation}
Within this procedure, those terms reducible to one- and two-site
correlation functions have been exactly evaluated. Because of the
translational symmetry, we have
\begin{equation}
\begin{array}{l}
\displaystyle I_{33\sigma}({\bf k}) \simeq \frac{1}{2}\left( \langle
n_{-\sigma} \rangle -p_{\sigma} \right)
 +\langle \eta^{\alpha\dagger}_{-\sigma}\eta^{\alpha}_{-\sigma} \rangle
 -\frac{1}{4}\langle \eta^{\dagger}_{-\sigma}\eta_{-\sigma} \rangle \\
 \displaystyle \hspace*{5mm} +\left(
  \langle \xi^{\alpha\dagger}_{-\sigma}\xi^{\alpha}_{-\sigma} \rangle
  -\frac{1}{4}\langle \xi^{\dagger}_{-\sigma}\xi_{-\sigma} \rangle \right)
  \left( \langle n_{\sigma} \rangle -\langle n_{-\sigma} \rangle \right)
  -\left( \langle \eta^{\alpha\dagger}_{-\sigma}\eta^{\alpha}_{-\sigma} \rangle
  -\frac{1}{4}\langle \eta^{\dagger}_{-\sigma}\eta_{-\sigma} \rangle \right)
  \left( \langle n_{\sigma} \rangle +\langle n_{-\sigma} \rangle \right) \\
 \displaystyle \hspace*{5mm} +\left( 1-\frac{1}{4} \right)
  \left( \langle \xi^{\dagger}_{-\sigma}c^{\alpha}_{-\sigma} \rangle
  \langle \xi^{\dagger}_{-\sigma}c^{\alpha}_{-\sigma} \rangle
  +\langle \eta^{\dagger}_{-\sigma}c^{\alpha}_{-\sigma} \rangle
  \langle \eta^{\dagger}_{-\sigma}c^{\alpha}_{-\sigma} \rangle
  +2\langle \xi^{\dagger}_{\sigma}c^{\alpha}_{\sigma} \rangle
  \langle c^{\dagger}_{-\sigma}c^{\alpha}_{-\sigma} \rangle \right) \\
\displaystyle \hspace*{5mm}+\alpha({\bf k}) \left[ \begin{array}{l}
 \displaystyle -2\Delta_{\sigma} \left\{ \langle n_{-\sigma} \rangle
 +\frac{1}{4}\left( 1-\langle n_{-\sigma} \rangle \right) \right\}
 -2\langle \xi^{\dagger}_{-\sigma}c^{\alpha}_{-\sigma} \rangle
  \langle \xi^{\alpha\dagger}_{-\sigma}c^{\alpha}_{-\sigma} \rangle
 +\frac{1}{2}\langle \xi^{\dagger}_{-\sigma}c^{\alpha}_{-\sigma} \rangle
  \langle \xi^{\dagger}_{-\sigma}\xi_{-\sigma} \rangle \\
 \displaystyle +2\langle \eta^{\dagger}_{-\sigma}c^{\alpha}_{-\sigma} \rangle
  \langle \eta^{\alpha\dagger}_{-\sigma}c^{\alpha}_{-\sigma} \rangle
 -\frac{1}{2}\langle \eta^{\dagger}_{-\sigma}c^{\alpha}_{-\sigma} \rangle
  \langle \eta^{\dagger}_{-\sigma}\eta_{-\sigma} \rangle
 -2\langle c^{\dagger}_{-\sigma}c^{\alpha}_{-\sigma} \rangle
  \langle \xi^{\alpha\dagger}_{\sigma}c^{\alpha}_{\sigma} \rangle
 +\frac{1}{2}\langle c^{\dagger}_{-\sigma}c^{\alpha}_{-\sigma} \rangle
  \langle \xi^{\dagger}_{\sigma}\xi_{\sigma} \rangle \\
 \displaystyle +2\langle \eta^{\dagger}_{\sigma}c^{\alpha}_{\sigma} \rangle
  \langle \eta^{\alpha\dagger}_{-\sigma}c^{\alpha}_{-\sigma} \rangle
 -\frac{1}{2}\langle \eta^{\dagger}_{\sigma}c^{\alpha}_{\sigma} \rangle
  \langle \eta^{\dagger}_{-\sigma}\eta_{-\sigma} \rangle
 -2\langle \eta^{\dagger}_{\sigma}c^{\alpha}_{\sigma} \rangle
  \langle \xi^{\alpha\dagger}_{-\sigma}c^{\alpha}_{-\sigma} \rangle
 +\frac{1}{2}\langle \eta^{\dagger}_{\sigma}c^{\alpha}_{\sigma} \rangle
  \langle \xi^{\dagger}_{-\sigma}\xi_{-\sigma} \rangle
\end{array} \right] \\
\displaystyle \hspace*{5mm}+\frac{1}{2}\beta_{1}({\bf k}) \left[
\begin{array}{l}
 \displaystyle \langle n_{-\sigma} \rangle \langle n_{-\sigma} \rangle
  \left( 1-\langle n_{-\sigma}\rangle \right)
 +2\langle n_{-\sigma} \rangle \langle c^{\dagger}_{-\sigma}c^{\alpha}_{-\sigma} \rangle
  \left( \langle c^{\dagger}_{-\sigma}c^{\alpha}_{-\sigma} \rangle
  +2\langle c^{\dagger}_{\sigma}c^{\alpha}_{\sigma} \rangle \right) \\
 \displaystyle -\langle c^{\dagger}_{-\sigma}c^{\beta_{1}}_{-\sigma} \rangle
  \langle c^{\dagger}_{-\sigma}c^{\beta_{1}}_{-\sigma} \rangle ( 1-\langle n_{-\sigma}\rangle )
 -\langle c^{\dagger}_{-\sigma}c^{\beta_{1}}_{-\sigma} \rangle
  \langle c^{\dagger}_{\sigma}c^{\beta_{1}}_{\sigma} \rangle ( 1-2\langle n_{\sigma}\rangle ) \\
 \displaystyle -2\left( \langle c^{\dagger}_{\sigma}c^{\alpha}_{\sigma} \rangle
  +\langle c^{\dagger}_{-\sigma}c^{\alpha}_{-\sigma} \rangle \right)^{2}
  \langle c^{\dagger}_{-\sigma}c^{\beta_{1}}_{-\sigma} \rangle
\end{array} \right] \\
\displaystyle \hspace*{5mm}+\frac{1}{4}\beta_{2}({\bf k}) \left[
\begin{array}{l}
 \displaystyle \langle n_{-\sigma} \rangle \langle n_{-\sigma} \rangle
  \left( 1-\langle n_{-\sigma}\rangle \right)
 +2\langle n_{-\sigma} \rangle \langle c^{\dagger}_{-\sigma}c^{\alpha}_{-\sigma} \rangle
  \left( \langle c^{\dagger}_{-\sigma}c^{\alpha}_{-\sigma} \rangle
  +2\langle c^{\dagger}_{\sigma}c^{\alpha}_{\sigma} \rangle \right) \\
 \displaystyle -\langle c^{\dagger}_{-\sigma}c^{\beta_{2}}_{-\sigma} \rangle
  \langle c^{\dagger}_{-\sigma}c^{\beta_{2}}_{-\sigma} \rangle ( 1-\langle n_{-\sigma}\rangle )
 -\langle c^{\dagger}_{-\sigma}c^{\beta_{2}}_{-\sigma} \rangle
  \langle c^{\dagger}_{\sigma}c^{\beta_{2}}_{\sigma} \rangle ( 1-2\langle n_{\sigma}\rangle ) \\
 \displaystyle -2\left( \langle c^{\dagger}_{\sigma}c^{\alpha}_{\sigma} \rangle
  +\langle c^{\dagger}_{-\sigma}c^{\alpha}_{-\sigma} \rangle \right)^{2}
  \langle c^{\dagger}_{-\sigma}c^{\beta_{2}}_{-\sigma} \rangle
\end{array} \right]
\end{array}
\end{equation}
\begin{equation}
\begin{array}{l}
\displaystyle I_{34\sigma}({\bf k}) \simeq \langle
\eta^{\alpha\dagger}_{-\sigma}\xi^{\alpha}_{-\sigma} \rangle
 \left( 1-2\langle n_{-\sigma} \rangle \right)
 +\left( 1-\frac{1}{4} \right) \left( 2\langle \eta^{\dagger}_{-\sigma}c^{\alpha}_{-\sigma} \rangle
  \langle c^{\dagger}_{-\sigma}\xi^{\alpha}_{-\sigma} \rangle
 +\langle c^{\dagger}_{\sigma}c^{\alpha}_{\sigma} \rangle
  \langle c^{\dagger}_{-\sigma}c^{\alpha}_{-\sigma} \rangle \right) \\
\displaystyle \hspace*{5mm} +\alpha({\bf k}) \left[ \begin{array}{l}
 \displaystyle -\langle c^{\dagger}_{-\sigma}c^{\alpha}_{-\sigma} \rangle
  \langle c^{\alpha\dagger}_{\sigma}c^{\alpha}_{\sigma} \rangle
  +\frac{1}{4}\langle c^{\dagger}_{-\sigma}c^{\alpha}_{-\sigma} \rangle \langle n_{\sigma} \rangle
  -\left( \langle \xi^{\alpha\dagger}_{-\sigma}\xi^{\alpha}_{-\sigma} \rangle
  -\langle \eta^{\alpha\dagger}_{-\sigma}\eta^{\alpha}_{-\sigma} \rangle \right) \\
 \displaystyle  \left( \langle \xi^{\dagger}_{\sigma}\xi^{\alpha}_{\sigma} \rangle
  -\langle \eta^{\dagger}_{\sigma}\eta^{\alpha}_{\sigma} \rangle \right)
 +\frac{1}{4} \left( \langle \xi^{\dagger}_{-\sigma}\xi_{-\sigma} \rangle
  -\langle \eta^{\dagger}_{-\sigma}\eta_{-\sigma} \rangle \right)
  \left( \langle \xi^{\dagger}_{\sigma}\xi^{\alpha}_{\sigma} \rangle
  -\langle \eta^{\dagger}_{\sigma}\eta^{\alpha}_{\sigma} \rangle \right)
\end{array} \right]
\end{array}
\end{equation}
\begin{equation}
\begin{array}{l}
\displaystyle I_{44\sigma}({\bf k}) \simeq \langle
\xi^{\alpha\dagger}_{-\sigma}\xi^{\alpha}_{-\sigma} \rangle
 +\frac{1}{4}\langle \eta^{\dagger}_{-\sigma}\eta_{-\sigma} \rangle \\
 \displaystyle \hspace*{5mm} -\left(
  \langle \xi^{\alpha\dagger}_{-\sigma}\xi^{\alpha}_{-\sigma} \rangle
  -\frac{1}{4}\langle \xi^{\dagger}_{-\sigma}\xi_{-\sigma} \rangle \right)
  \left( \langle n_{\sigma} \rangle +\langle n_{-\sigma} \rangle \right)
  +\left( \langle \eta^{\alpha\dagger}_{-\sigma}\eta^{\alpha}_{-\sigma} \rangle
  -\frac{1}{4}\langle \eta^{\dagger}_{-\sigma}\eta_{-\sigma} \rangle \right)
  \left( \langle n_{\sigma} \rangle -\langle n_{-\sigma} \rangle \right) \\
 \displaystyle \hspace*{5mm} +\left( 1-\frac{1}{4} \right)
  \left( \langle \xi^{\dagger}_{-\sigma}c^{\alpha}_{-\sigma} \rangle
  \langle \xi^{\dagger}_{-\sigma}c^{\alpha}_{-\sigma} \rangle
  +\langle \eta^{\dagger}_{-\sigma}c^{\alpha}_{-\sigma} \rangle
  \langle \eta^{\dagger}_{-\sigma}c^{\alpha}_{-\sigma} \rangle
  +2\langle \eta^{\dagger}_{\sigma}c^{\alpha}_{\sigma} \rangle
  \langle c^{\dagger}_{-\sigma}c^{\alpha}_{-\sigma} \rangle \right) \\
\displaystyle \hspace*{5mm}+\alpha({\bf k}) \left[ \begin{array}{l}
 \displaystyle 2\left( 1-\frac{1}{4}\right) \Delta_{\sigma}\langle n_{-\sigma} \rangle
 +2\langle \xi^{\dagger}_{-\sigma}c^{\alpha}_{-\sigma} \rangle
  \langle \xi^{\alpha\dagger}_{-\sigma}c^{\alpha}_{-\sigma} \rangle
 -\frac{1}{2}\langle \xi^{\dagger}_{-\sigma}c^{\alpha}_{-\sigma} \rangle
  \langle \xi^{\dagger}_{-\sigma}\xi_{-\sigma} \rangle \\
 \displaystyle -2\langle \eta^{\dagger}_{-\sigma}c^{\alpha}_{-\sigma} \rangle
  \langle \eta^{\alpha\dagger}_{-\sigma}c^{\alpha}_{-\sigma} \rangle
 +\frac{1}{2}\langle \eta^{\dagger}_{-\sigma}c^{\alpha}_{-\sigma} \rangle
  \langle \eta^{\dagger}_{-\sigma}\eta_{-\sigma} \rangle
 -2\langle c^{\dagger}_{-\sigma}c^{\alpha}_{-\sigma} \rangle
  \langle \eta^{\alpha\dagger}_{\sigma}c^{\alpha}_{\sigma} \rangle
 +\frac{1}{2}\langle c^{\dagger}_{-\sigma}c^{\alpha}_{-\sigma} \rangle
  \langle \eta^{\dagger}_{\sigma}\eta_{\sigma} \rangle \\
 \displaystyle +2\langle \xi^{\dagger}_{\sigma}c^{\alpha}_{\sigma} \rangle
  \langle \xi^{\alpha\dagger}_{-\sigma}c^{\alpha}_{-\sigma} \rangle
 -\frac{1}{2}\langle \xi^{\dagger}_{\sigma}c^{\alpha}_{\sigma} \rangle
  \langle \xi^{\dagger}_{-\sigma}\xi_{-\sigma} \rangle
 -2\langle \xi^{\dagger}_{\sigma}c^{\alpha}_{\sigma} \rangle
  \langle \eta^{\alpha\dagger}_{-\sigma}c^{\alpha}_{-\sigma} \rangle
 +\frac{1}{2}\langle \xi^{\dagger}_{\sigma}c^{\alpha}_{\sigma} \rangle
  \langle \eta^{\dagger}_{-\sigma}\eta_{-\sigma} \rangle
\end{array} \right] \\
\displaystyle \hspace*{5mm}+\frac{1}{2}\beta_{1}({\bf k}) \left[
\begin{array}{l}
 \displaystyle \langle n_{-\sigma} \rangle \langle n_{-\sigma} \rangle \langle n_{-\sigma} \rangle
 -2\langle n_{-\sigma} \rangle \langle c^{\dagger}_{-\sigma}c^{\alpha}_{-\sigma} \rangle
  \left( \langle c^{\dagger}_{-\sigma}c^{\alpha}_{-\sigma} \rangle
  +2\langle c^{\dagger}_{\sigma}c^{\alpha}_{\sigma} \rangle \right) \\
 \displaystyle -\langle c^{\dagger}_{-\sigma}c^{\beta_{1}}_{-\sigma} \rangle
  \langle c^{\dagger}_{-\sigma}c^{\beta_{1}}_{-\sigma} \rangle \langle n_{-\sigma}\rangle
 +\langle c^{\dagger}_{-\sigma}c^{\beta_{1}}_{-\sigma} \rangle
  \langle c^{\dagger}_{\sigma}c^{\beta_{1}}_{\sigma} \rangle ( 1-2\langle n_{\sigma}\rangle ) \\
 \displaystyle +2\left( \langle c^{\dagger}_{\sigma}c^{\alpha}_{\sigma} \rangle
  +\langle c^{\dagger}_{-\sigma}c^{\alpha}_{-\sigma} \rangle \right)^{2}
  \langle c^{\dagger}_{-\sigma}c^{\beta_{1}}_{-\sigma} \rangle
\end{array} \right] \\
\displaystyle \hspace*{5mm}+\frac{1}{4}\beta_{2}({\bf k}) \left[
\begin{array}{l}
 \displaystyle \langle n_{-\sigma} \rangle \langle n_{-\sigma} \rangle \langle n_{-\sigma} \rangle
 -2\langle n_{-\sigma} \rangle \langle c^{\dagger}_{-\sigma}c^{\alpha}_{-\sigma} \rangle
  \left( \langle c^{\dagger}_{-\sigma}c^{\alpha}_{-\sigma} \rangle
  +2\langle c^{\dagger}_{\sigma}c^{\alpha}_{\sigma} \rangle \right) \\
 \displaystyle -\langle c^{\dagger}_{-\sigma}c^{\beta_{2}}_{-\sigma} \rangle
  \langle c^{\dagger}_{-\sigma}c^{\beta_{2}}_{-\sigma} \rangle \langle n_{-\sigma}\rangle
 +\langle c^{\dagger}_{-\sigma}c^{\beta_{2}}_{-\sigma} \rangle
  \langle c^{\dagger}_{\sigma}c^{\beta_{2}}_{\sigma} \rangle ( 1-2\langle n_{\sigma}\rangle ) \\
 \displaystyle +2\left( \langle c^{\dagger}_{\sigma}c^{\alpha}_{\sigma} \rangle
  +\langle c^{\dagger}_{-\sigma}c^{\alpha}_{-\sigma} \rangle \right)^{2}
  \langle c^{\dagger}_{-\sigma}c^{\beta_{2}}_{-\sigma} \rangle
\end{array} \right]
\end{array}
\end{equation}
where
\begin{equation}
\begin{array}{l}
\displaystyle \alpha({\bf k})=\frac{1}{2}\left\{ \cos k_x a_x +\cos k_y a_y \right\} \\
\displaystyle \beta_{1}({\bf k})=\cos k_x a_x \cos k_y a_y \\
\displaystyle \beta_{2}({\bf k})=\frac{1}{2}\left\{ \cos 2k_x a_x
+\cos 2k_y a_y \right\}
\end{array}
\end{equation}
with
\begin{equation}
\alpha^{2}({\bf k})=\frac{1}{4}+\frac{1}{2}\beta_{1}({\bf
k})+\frac{1}{4}\beta_{2}({\bf k}).
\end{equation}
Parameters $p_{\sigma}$ and $\Delta_{\sigma}$ are defined in the
text.

$\langle A(i)B^{\beta_1}(i)\rangle$ and $\langle
A(i)B^{\beta_2}(i)\rangle$ contain next-nearest-neighbor-site
correlation functions along the diagonal and the main directions,
respectively. If we assume that those correlation functions have
same value, that is, we use the spherical approximation, we can
simplify the momentum dependence of $I$
\begin{equation}
\begin{array}{l}
\displaystyle I_{33\sigma}({\bf k}) \simeq \frac{1}{2}\left( \langle
n_{-\sigma} \rangle -p_{\sigma} \right)
 +\langle \eta^{\alpha\dagger}_{-\sigma}\eta^{\alpha}_{-\sigma} \rangle
 -\frac{1}{4}\langle \eta^{\dagger}_{-\sigma}\eta_{-\sigma} \rangle \\
 \displaystyle \hspace*{5mm} +\left(
  \langle \xi^{\alpha\dagger}_{-\sigma}\xi^{\alpha}_{-\sigma} \rangle
  -\frac{1}{4}\langle \xi^{\dagger}_{-\sigma}\xi_{-\sigma} \rangle \right)
  \left( \langle n_{\sigma} \rangle -\langle n_{-\sigma} \rangle \right)
  -\left( \langle \eta^{\alpha\dagger}_{-\sigma}\eta^{\alpha}_{-\sigma} \rangle
  -\frac{1}{4}\langle \eta^{\dagger}_{-\sigma}\eta_{-\sigma} \rangle \right)
  \left( \langle n_{\sigma} \rangle +\langle n_{-\sigma} \rangle \right) \\
 \displaystyle \hspace*{5mm} +\left( 1-\frac{1}{4} \right)
  \left( \langle \xi^{\dagger}_{-\sigma}c^{\alpha}_{-\sigma} \rangle
  \langle \xi^{\dagger}_{-\sigma}c^{\alpha}_{-\sigma} \rangle
  +\langle \eta^{\dagger}_{-\sigma}c^{\alpha}_{-\sigma} \rangle
  \langle \eta^{\dagger}_{-\sigma}c^{\alpha}_{-\sigma} \rangle
  +2\langle \xi^{\dagger}_{\sigma}c^{\alpha}_{\sigma} \rangle
  \langle c^{\dagger}_{-\sigma}c^{\alpha}_{-\sigma} \rangle \right) \\
\displaystyle \hspace*{5mm}+\alpha({\bf k}) \left[ \begin{array}{l}
 \displaystyle -2\Delta_{\sigma} \left\{ \langle n_{-\sigma} \rangle
 +\frac{1}{4}\left( 1-\langle n_{-\sigma} \rangle \right) \right\}
 -2\langle \xi^{\dagger}_{-\sigma}c^{\alpha}_{-\sigma} \rangle
  \langle \xi^{\alpha\dagger}_{-\sigma}c^{\alpha}_{-\sigma} \rangle
 +\frac{1}{2}\langle \xi^{\dagger}_{-\sigma}c^{\alpha}_{-\sigma} \rangle
  \langle \xi^{\dagger}_{-\sigma}\xi_{-\sigma} \rangle \\
 \displaystyle +2\langle \eta^{\dagger}_{-\sigma}c^{\alpha}_{-\sigma} \rangle
  \langle \eta^{\alpha\dagger}_{-\sigma}c^{\alpha}_{-\sigma} \rangle
 -\frac{1}{2}\langle \eta^{\dagger}_{-\sigma}c^{\alpha}_{-\sigma} \rangle
  \langle \eta^{\dagger}_{-\sigma}\eta_{-\sigma} \rangle
 -2\langle c^{\dagger}_{-\sigma}c^{\alpha}_{-\sigma} \rangle
  \langle \xi^{\alpha\dagger}_{\sigma}c^{\alpha}_{\sigma} \rangle
 +\frac{1}{2}\langle c^{\dagger}_{-\sigma}c^{\alpha}_{-\sigma} \rangle
  \langle \xi^{\dagger}_{\sigma}\xi_{\sigma} \rangle \\
 \displaystyle +2\langle \eta^{\dagger}_{\sigma}c^{\alpha}_{\sigma} \rangle
  \langle \eta^{\alpha\dagger}_{-\sigma}c^{\alpha}_{-\sigma} \rangle
 -\frac{1}{2}\langle \eta^{\dagger}_{\sigma}c^{\alpha}_{\sigma} \rangle
  \langle \eta^{\dagger}_{-\sigma}\eta_{-\sigma} \rangle
 -2\langle \eta^{\dagger}_{\sigma}c^{\alpha}_{\sigma} \rangle
  \langle \xi^{\alpha\dagger}_{-\sigma}c^{\alpha}_{-\sigma} \rangle
 +\frac{1}{2}\langle \eta^{\dagger}_{\sigma}c^{\alpha}_{\sigma} \rangle
  \langle \xi^{\dagger}_{-\sigma}\xi_{-\sigma} \rangle
\end{array} \right] \\
\displaystyle \hspace*{5mm}+\beta({\bf k}) \left[ \begin{array}{l}
 \displaystyle \langle n_{-\sigma} \rangle \langle n_{-\sigma} \rangle
  \left( 1-\langle n_{-\sigma}\rangle \right)
 +2\langle n_{-\sigma} \rangle \langle c^{\dagger}_{-\sigma}c^{\alpha}_{-\sigma} \rangle
  \left( \langle c^{\dagger}_{-\sigma}c^{\alpha}_{-\sigma} \rangle
  +2\langle c^{\dagger}_{\sigma}c^{\alpha}_{\sigma} \rangle \right) \\
 \displaystyle -\langle c^{\dagger}_{-\sigma}c^{\beta}_{-\sigma} \rangle
  \langle c^{\dagger}_{-\sigma}c^{\beta}_{-\sigma} \rangle ( 1-\langle n_{-\sigma}\rangle )
 -\langle c^{\dagger}_{-\sigma}c^{\beta}_{-\sigma} \rangle
  \langle c^{\dagger}_{\sigma}c^{\beta}_{\sigma} \rangle ( 1-2\langle n_{\sigma}\rangle ) \\
 \displaystyle -2\left( \langle c^{\dagger}_{\sigma}c^{\alpha}_{\sigma} \rangle
  +\langle c^{\dagger}_{-\sigma}c^{\alpha}_{-\sigma} \rangle \right)^{2}
  \langle c^{\dagger}_{-\sigma}c^{\beta}_{-\sigma} \rangle
\end{array} \right]
\end{array}
\end{equation}
\begin{equation}
\begin{array}{l}
\displaystyle I_{44\sigma}({\bf k}) \simeq \langle
\xi^{\alpha\dagger}_{-\sigma}\xi^{\alpha}_{-\sigma} \rangle
 +\frac{1}{4}\langle \eta^{\dagger}_{-\sigma}\eta_{-\sigma} \rangle \\
 \displaystyle \hspace*{5mm} -\left(
  \langle \xi^{\alpha\dagger}_{-\sigma}\xi^{\alpha}_{-\sigma} \rangle
  -\frac{1}{4}\langle \xi^{\dagger}_{-\sigma}\xi_{-\sigma} \rangle \right)
  \left( \langle n_{\sigma} \rangle +\langle n_{-\sigma} \rangle \right)
  +\left( \langle \eta^{\alpha\dagger}_{-\sigma}\eta^{\alpha}_{-\sigma} \rangle
  -\frac{1}{4}\langle \eta^{\dagger}_{-\sigma}\eta_{-\sigma} \rangle \right)
  \left( \langle n_{\sigma} \rangle -\langle n_{-\sigma} \rangle \right) \\
 \displaystyle \hspace*{5mm} +\left( 1-\frac{1}{4} \right)
  \left( \langle \xi^{\dagger}_{-\sigma}c^{\alpha}_{-\sigma} \rangle
  \langle \xi^{\dagger}_{-\sigma}c^{\alpha}_{-\sigma} \rangle
  +\langle \eta^{\dagger}_{-\sigma}c^{\alpha}_{-\sigma} \rangle
  \langle \eta^{\dagger}_{-\sigma}c^{\alpha}_{-\sigma} \rangle
  +2\langle \eta^{\dagger}_{\sigma}c^{\alpha}_{\sigma} \rangle
  \langle c^{\dagger}_{-\sigma}c^{\alpha}_{-\sigma} \rangle \right) \\
\displaystyle \hspace*{5mm}+\alpha({\bf k}) \left[ \begin{array}{l}
 \displaystyle 2\left( 1-\frac{1}{4}\right) \Delta_{\sigma}\langle n_{-\sigma} \rangle
 +2\langle \xi^{\dagger}_{-\sigma}c^{\alpha}_{-\sigma} \rangle
  \langle \xi^{\alpha\dagger}_{-\sigma}c^{\alpha}_{-\sigma} \rangle
 -\frac{1}{2}\langle \xi^{\dagger}_{-\sigma}c^{\alpha}_{-\sigma} \rangle
  \langle \xi^{\dagger}_{-\sigma}\xi_{-\sigma} \rangle \\
 \displaystyle -2\langle \eta^{\dagger}_{-\sigma}c^{\alpha}_{-\sigma} \rangle
  \langle \eta^{\alpha\dagger}_{-\sigma}c^{\alpha}_{-\sigma} \rangle
 +\frac{1}{2}\langle \eta^{\dagger}_{-\sigma}c^{\alpha}_{-\sigma} \rangle
  \langle \eta^{\dagger}_{-\sigma}\eta_{-\sigma} \rangle
 -2\langle c^{\dagger}_{-\sigma}c^{\alpha}_{-\sigma} \rangle
  \langle \eta^{\alpha\dagger}_{\sigma}c^{\alpha}_{\sigma} \rangle
 +\frac{1}{2}\langle c^{\dagger}_{-\sigma}c^{\alpha}_{-\sigma} \rangle
  \langle \eta^{\dagger}_{\sigma}\eta_{\sigma} \rangle \\
 \displaystyle +2\langle \xi^{\dagger}_{\sigma}c^{\alpha}_{\sigma} \rangle
  \langle \xi^{\alpha\dagger}_{-\sigma}c^{\alpha}_{-\sigma} \rangle
 -\frac{1}{2}\langle \xi^{\dagger}_{\sigma}c^{\alpha}_{\sigma} \rangle
  \langle \xi^{\dagger}_{-\sigma}\xi_{-\sigma} \rangle
 -2\langle \xi^{\dagger}_{\sigma}c^{\alpha}_{\sigma} \rangle
  \langle \eta^{\alpha\dagger}_{-\sigma}c^{\alpha}_{-\sigma} \rangle
 +\frac{1}{2}\langle \xi^{\dagger}_{\sigma}c^{\alpha}_{\sigma} \rangle
  \langle \eta^{\dagger}_{-\sigma}\eta_{-\sigma} \rangle
\end{array} \right] \\
\displaystyle \hspace*{5mm}+\beta({\bf k}) \left[ \begin{array}{l}
 \displaystyle \langle n_{-\sigma} \rangle \langle n_{-\sigma} \rangle \langle n_{-\sigma} \rangle
 -2\langle n_{-\sigma} \rangle \langle c^{\dagger}_{-\sigma}c^{\alpha}_{-\sigma} \rangle
  \left( \langle c^{\dagger}_{-\sigma}c^{\alpha}_{-\sigma} \rangle
  +2\langle c^{\dagger}_{\sigma}c^{\alpha}_{\sigma} \rangle \right) \\
 \displaystyle -\langle c^{\dagger}_{-\sigma}c^{\beta}_{-\sigma} \rangle
  \langle c^{\dagger}_{-\sigma}c^{\beta}_{-\sigma} \rangle \langle n_{-\sigma}\rangle
 +\langle c^{\dagger}_{-\sigma}c^{\beta}_{-\sigma} \rangle
  \langle c^{\dagger}_{\sigma}c^{\beta}_{\sigma} \rangle ( 1-2\langle n_{\sigma}\rangle ) \\
 \displaystyle +2\left( \langle c^{\dagger}_{\sigma}c^{\alpha}_{\sigma} \rangle
  +\langle c^{\dagger}_{-\sigma}c^{\alpha}_{-\sigma} \rangle \right)^{2}
  \langle c^{\dagger}_{-\sigma}c^{\beta}_{-\sigma} \rangle
\end{array} \right]
\end{array}
\end{equation}
with
\begin{equation}
\beta({\bf k})=\frac{1}{2}\beta_{1}({\bf
k})+\frac{1}{4}\beta_{2}({\bf k})
 =\alpha^{2}({\bf k})-\frac{1}{4}.
\end{equation}
We have checked that this assumption doesn't produce any difference
as regards the results reported in the present paper. We have used
this approximation in order to simplify  the momentum integration.

\end{widetext}


\end{document}